\documentclass[preprint,showpacs,groupedaddress,amsmath,amssymb]{revtex4-1}  
\usepackage{graphicx,color}
\usepackage{multirow}
\begin{document}
\title{ Electronic structure of stacking faults in hexagonal graphite }
\author{ M. Taut, K. Koepernik, and M. Richter   }
\affiliation{\small IFW Dresden, PO Box 270116, D-01171 Dresden, Germany}
\date{{\small (\today)}}

\begin{abstract} 
We present results of  self-consistent, full-potential   
electronic structure calculations for 
slabs of hexagonal graphite with stacking faults and 
for slabs with one displaced surface layer. 
There are two types of stacking faults, 
which differ qualitatively in their chemical  bonding picture. 
We find, that both types induce localized 
interface bands near the symmetry line K-M 
in the Brillouin zone  and 
a related peak in the local density of states (LDOS) very close to
the Fermi energy, which
should give rise to a dominating contribution of the interface bands 
to the local conductivity at the stacking faults. 
In contrast, a clean surface does not host any 
surface bands in the energy range of the $\pi$ and $\sigma$ bands,
and  the LDOS  near the surface is even depleted. 
On the other hand, displacement of even one single surface layer 
induces a surface band near K-M. 
A special role play $p_z$-bonded dimers 
(directed perpendicular to the layers) in the vicinity of 
one type of stacking faults. 
They  produce a half-filled pair of  interface states / interface resonances.
The formation energy of both types of stacking faults and the surface energy 
are estimated. 
\end{abstract}


\maketitle


\section{ Introduction}

Graphitic systems gained a renewed interest 
(after the  intercalation, fullerene,   and nanotube rushs) 
following the invention  of techniques to produce thin two-dimensional systems 
in the form of thin slabs or even mono-layers (for recent review see e.g. 
Refs.~\onlinecite{neto09, abergel10, peres10, goerbig11}). 
There is, however, another form of two-dimensionality,
namely stacking faults in bulk hexagonal (AB) graphite with related 
interface states. 
These states are interesting, because 
they have been suspected to play a role in the observed  integer 
quantum Hall effect \cite{kempa06,guinea08} in bulk graphite.
As we will show, they may also contribute considerably to  the 
electronic conductivity.

\begin{figure}[h]
\includegraphics[width=0.5\textwidth]{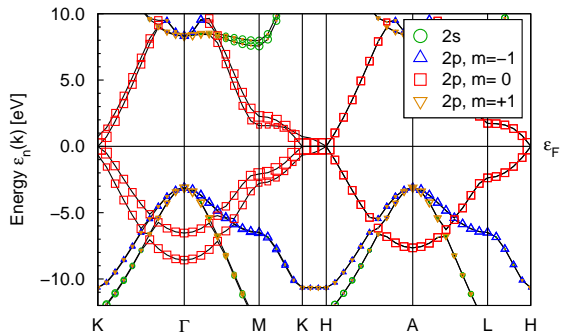}
\vspace{-1cm}
\caption{(color online) Band weights of the orbitals of the second shell
 at a {\bf single atom} in hexagonal bulk graphite. 
The figure for the {\bf chain atoms} (not shown) differs mainly  near 
the line K-H around the Fermi energy (see the zoom in Fig.~\ref{fig:BW-at-EF-AB}).}
\label{fig:BW-all-lines-AB} 
\end{figure}

The electronic states in graphitic piles can be grouped into 
$\pi$ and $\sigma$ states.
The $\pi$ bands are mainly formed by  $p_z$ orbitals, they
are responsible for the electronic structure
around the Fermi energy $\varepsilon_F$.
The $\sigma$ bands with a total band width of about 40 eV are formed by
the three $sp_{(2)}$ linear combinations of $s, p_x$, and $ p_y$ orbitals
and lie more than 3 eV below and more than 8 eV above $\varepsilon_F$ 
(see Fig.~\ref{fig:BW-all-lines-AB}).
It is mainly the  $\sigma$ bands which stabilize the honeycomb structure 
of the graphene layers,
while the $\pi$ bands account for their intriguing physical properties.

The importance of the electronic structure of stacking 
faults in graphite results from 
their abundance in all kinds of samples and from the fact that they host 
localized interface bands~\cite{guinea08, koshino13}. 
The large probability of their occurrence  has its origin in 
the small energy difference between different stacking sequences, 
which for its part  can be traced back to the 
large difference between intra-layer 
and inter-layer  overlap integrals. 
The largest interlayer overlap integral between $p_z$ orbitals, $\gamma_1$,
is $\approx$ 0.4 eV,
and the largest  intra-layer overlap integral between $p_z$ orbitals,
$\gamma_0$, is $\approx$ 2.6 eV~\cite{brandt88,grueneis08}. 
The  overlap integrals between 
$sp_{(2)}$ orbitals are not included in the common tight-bonding models.
It follows from the large $\sigma$ gap 
and bandwidth, that they are much larger than $\gamma_0$. 

\begin{figure}[h]
\centering
\includegraphics[width=0.5\textwidth]{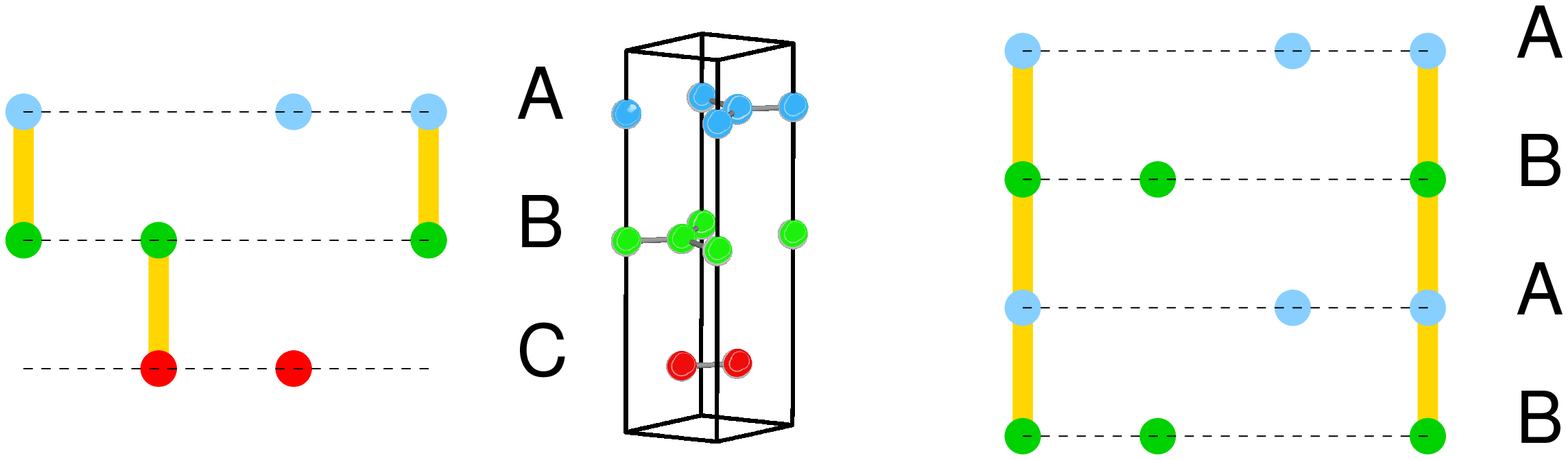}
\caption{ (color online)
The three basic layers A (blue), B (green), and C (red)
in graphitic stacks within a hexagonal unit cell.
Left: atoms on a plane in a hexagonal  unit cell of rhombohedral 
(ABC) graphite,
 and right: in two unit cells of hexagonal (Bernal) stacking. 
Saturated $p_z$ bonds are indicated by
yellow perpendicular bars. }
\label{fig:atoms-ABC}
\end{figure}

Fig.~\ref{fig:atoms-ABC}
shows the  three possible highly-symmetric relative locations 
of graphene layers within a hexagonal unit cell. 
In constructing stacking faults, we consider slabs of AB-stacks
(which would form hexagonal graphite, if the slab was infinite), 
followed by a C-layer, 
whereby the C-layer is already part of  the subsequent CA- or CB-stack. 
In this way, two kinds of stacking faults are generated (see Sect.~IV).
Generally, we rule out the neighborhood of two identical layers,  because of
 their large contribution to the total energy 
(see \cite{gonze94} and our results below).
Such a slab with a stacking fault is either periodically repeated 
(without any surface), or surrounded by  slabs of vacuum and
 then periodically repeated. 
In the latter case  we can study interfaces and surfaces in the same 
calculation. In case of periodic repetition without vacuum 
we have the advantages that the unit cell can be chosen to have 
a higher symmetry than a slab, and that the large surface energy does not mask the small 
total energy differences between the two types of stacking faults. 
Therefore, total energies of stacking faults 
 were calculated in the geometry without surfaces, 
but one-electron spectra and surface energies in slabs with surfaces.
Additionally, we considere pure AB-slabs with surfaces (see Sect.~III) and 
with a single displaced surface layer of type C on either side (see Sect.~VI). 
The width of the vacuum layer was chosen to be 6 interlayer distances 
throughout this paper.

Two previous works on stacking faults \cite{guinea08, koshino13} considered two
phenomenological overlap integrals ($\gamma_0$ and $\gamma_1$)
within a mostly analytical model calculation. 
While this model correctly predicts the existence (not the details)
of interface bands at one of the two possible
stacking faults in hexagonal graphite, it cannot describe
the band structure around $\varepsilon_F$ of hexagonal (Bernal) or rhombohedral
bulk graphite even qualitatively (see Ref.~\onlinecite{koshino13} and
references therein).
Former self-consistent slab calculations on graphitic slabs 
can be found in Refs.~\cite{latil06,aoki07,min07,zhang10,xiao11}. 

In this paper we present results from self-consistent full-potential
local-orbital calculations using the FPLO-package \cite{fplo,koepernik99a}.
All calculations use density functional theory (DFT). {\em Total energies }
of stacking faults
are calculated within the local density approximation (LDA) 
(PW92 \cite{PW92}).
The LDA has been chosen for the
total energy, because it benefits from a cancellation
between the over-binding (characteristic for the LDA) and the neglected
van-der-Waals interaction. 
By minimizing the total energy
the LDA provides   rather precise
values for the lattice constants of hexagonal graphite  (see \cite{gonze94} and our results below).
For the {\em one-electron spectra} we used the
generalized gradient approximation (GGA) 
(PBE96 \cite{PBE96}), because the LDA fails to
reproduce the characteristic four-leg structure of the electron pocket in
hexagonal bulk graphite (see Sect.~II).
The GGA, on the other hand, which in many cases provides 
more accurate lattice constants than the LDA,
fails to describe the binding between graphene layers 
(see also \cite{ooi06}).
All calculations on one-electron spectra were done with the experimental 
C-C distance within the layers of 1.42 \AA {}  
and an interlayer distance of 3.33 \AA{}~\cite{aoki07}. 
The lattice constants from total energy calculations are separately discussed 
in the text.

Band weights $W_{\nu}({\bf k},n) = |\langle \nu | {\bf k},n\rangle |^2 $
 represent the size of the contribution of 
local orbital 
$|\nu\rangle$ to the Bloch wave function $|{\bf k},n\rangle$. 
The sum of all orbital weights is normalized  to unity  
\begin{equation}
\sum_\nu W_\nu({\bf k},n)=1. 
\label{sumrule}
\end{equation}
In the plots with band weights, 
the local orbitals $|\nu\rangle$ are represented by the form and color
of symbols 
at the energy bands, and the band weights by the size of the symbols. 
Because for {\em bulk states} the band weights converge to constants 
$W_{\nu}^{bulk}({\bf k},n)$ away from surfaces/interfaces, 
it follows from the sum rule (\ref{sumrule}) that $W_{\nu}^{bulk}$
scales like $1/N_{\parallel} N_{\perp}$, 
where $N_{\parallel}$ and $N_{\perp}$ are the number of atoms in a layer and the number of layers in a slab, respectively. 
For {\em surface / interface states } the band weights converge to zero, 
and consequently the band weights scale like $1/N_{\parallel}$. 
This means that for bulk states the band weights at each site are decreasing
 with growing slab thickness, whereas for surface / interface states they
 become independent of the thickness. For thick slabs the largest 
band weights for 
surface / interface states (in the vicinity of the surface / interface)
 are therefore much larger than for bulk states. \\
The local density of states (LDOS) is defined as 
\begin{equation}
D_\nu(E) = 2 \sum_{{\bf k},n} W_{\nu}({\bf k},n) \;\delta(E-\varepsilon_n({\bf k})) 
\label{defLDOS}
\end{equation}
and integrates to 2 electrons. 
All $\bf k$ space integrations were done with 
the tetrahedron method \cite{tetrahedron1,tetrahedron2}.


\section{Bulk graphite}

\subsection{Band structure}
Despite the fact that hexagonal bulk graphite has already been the subject 
of numerous works (see e.g. reviews on older work \cite{kelly81, brandt88}, 
 and the more  recent self-consistent calculations using LDA, GGA, and GW 
\cite{gonze91, gonze94, ooi06, grueneis08}),
 we have to say a few words 
about it.
This is because we want to demonstrate the results of our approach 
in  the bulk limit, and to introduce the 
projected bulk  band structure (PBBS) of hexagonal graphite. 
If not otherwise indicated, the {\bf k}-mesh
for the self-consistent calculation on bulk hexagonal graphite comprises 500x500x100
inequivalent  points distributed equidistantly over the Brillouin zone (BZ).

\begin{figure}[h]
\includegraphics[width=0.5\textwidth]{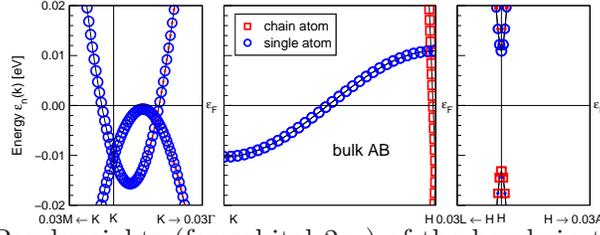}
\vspace{-1cm}
\caption{(color online) Band weights (for orbital $2 p_z$) 
of the bands in the vicinity of the 
line K-H at the two
 Wyckoff positions (single and chain atom) in hexagonal bulk graphite. 
The scale of the lines on M-K-$\Gamma$ and L-H-A is the same, 
but differs from the scale of K-H. 
Only those 3\% of the lines M-K, K-$\Gamma$, L-H, and H-A are shown, which lie  
 closest to the points K or H. As to the notation of the end points: 
the point K $\rightarrow$ 0.03 M lies on the line K-M, shifted by 3\% from point K 
toward point M, etc.
}
\label{fig:BW-at-EF-AB}    
\end{figure}

\begin{figure}[h]
\centering
\includegraphics[width=0.5\textwidth]{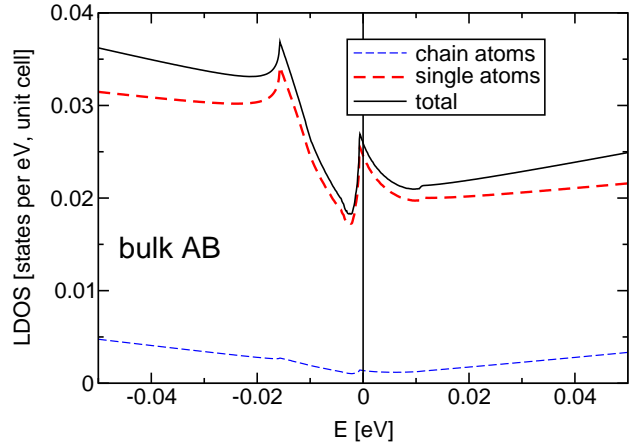}
\caption{(color online) LDOS of hexagonal bulk graphite
(projection on all orbitals per site). The Fermi level in all 
LDOS plots is at zero energy. 
The LDOS has been calculated with a special tetrahedron mesh 
located in a restricted  part of the BZ around the line K-H including 150x150x150 points. }
\label{fig:bulk-LDOS-AB}  
\end{figure}

Fig.~\ref{fig:BW-all-lines-AB} gives an overview over 
the bulk band structure in GGA  showing that 
$\varepsilon_F$ lies in a gap of the $\sigma$-bands of width \mbox{$\approx$ 11 eV}  
 and that all the Fermi surface physics comes from $p_z$ bands. 
Perpendicular to the layers,
bulk hexagonal graphite consists of linear chains of 
atoms with overlapping $p_z$ orbitals and 
single atoms (monomers)  with dangling $p_z$ bonds
(see  Fig.~\ref{fig:atoms-ABC}).
Fig.~\ref{fig:BW-at-EF-AB} and Fig.~\ref{fig:bulk-LDOS-AB}  show 
that it is mainly the single atoms 
which carry the states around $\varepsilon_F$.
In particular, the states of the degenerate very flat band with 20 meV dispersion 
between K and H carry only weights from the single atoms (Fig.~\ref{fig:BW-at-EF-AB}). 
The other bands with about 1 eV dispersion between K and H carry only weights 
from the chain atoms. The small dispersion of the single-atom bands is due
 to their dangling bonds. 
The two peaks in the LDOS near $\varepsilon_F$ in Fig.~\ref{fig:bulk-LDOS-AB}
are produced by the two extrema 
seen in the left panel of Fig.\ref{fig:BW-at-EF-AB}, 
which refers to the central plane of the BZ at $k_z$=0. 

Our Fermi surface (FS) in GGA shows the generally accepted four-leg 
topology of the {\em majority electron- and hole pockets}, which are
located correctly within the BZ. 
The only qualitative difference in our LDA results (not shown) is that 
the maximum of the downward bent 
parabola in the left panel of Fig.~\ref{fig:BW-at-EF-AB} 
 lies slightly (by 0.5 meV) above $\varepsilon_F$. This tiny shift has the consequence that 
the electron pocket decomposes into 4 pieces, just as reported in the 
LDA approach by Ref.~\cite{gonze91}.  Therefore, we used the 
GGA for the calculation of all one-particle spectra. 
  
In the literature there has been a lengthy debate about the number
and location of tiny {\em minority pockets}, 
which depend sensitively on  the sign and 
size of the small overlap integrals  (mainly $\gamma_6$) of the 
Slonzewski-Weiss-McClure (SWM) model \cite{SWM1,SWM2} 
(see review \cite{brandt88}) and,  
from the experimental side, on the characteristics of the samples. 
Even recently appeared papers on the interpretation of the
de Haas - van Alphen data \cite{kopelevich04,sugawa06} indicating that 
the matter is still in discussion. 
With our DFT calculations we did not find any minority pockets, 
which seems to be a general trait 
of self-consistent DFT and GW  calculations \cite{gonze91, grueneis08}. 
This problem can be fixed by introducing artificial doping \cite{grueneis08},
but we did not take any measures 
in order to make the FS agree with the SWM model
in the issue of the tiny minority pockets.


\subsection{Projected bulk band structure}
\begin{figure}[h]
\vspace{1cm}
\includegraphics[width=0.5\textwidth]{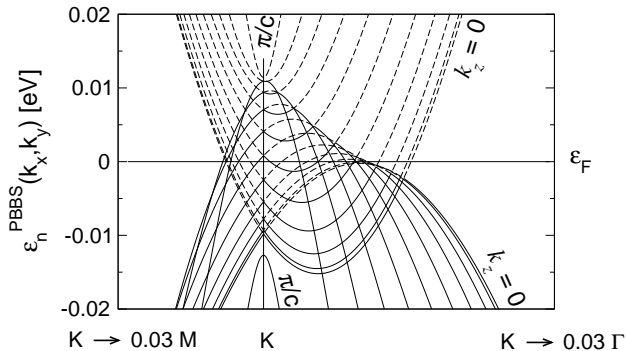}
\caption{
Bulk band structure of (AB) graphite projected onto a plane
parallel to the layers for 11 equidistant $k_z$ values. 
Only 3 \% of the  lines M-K and K-$\Gamma$
in the ${\bf k}_{\parallel}$-space,
centered around symmetry point K, are shown. 
Full and dashed lines denote the valence and conduction band, respectively.  }
\label{fig:projBS} 
\end{figure}

The PBBS is a useful tool to separate surface- or interface bands   
of thick slabs from bulk bands 
without investigating all slab wave-functions in detail, 
but just by locating their energy relative to the PBBS. 
The PBBS is defined as follows:
Assume that the slabs to be investigated extend in the $x-y$ plane and calculate 
the bulk band structure (BBS) 
$\varepsilon^{\rm bulk}_n(k_x,k_y,k_z)$ with macroscopic periodic
boundary condition in all 3 dimensions.  
Then plot $\varepsilon_n^{\rm bulk}(k_x,k_y)$ for all $n$ and for
a quasi-continuous set of $k_z$ values: 
\begin{equation}
\varepsilon^{\rm PBBS}_n (k_x, k_y) \stackrel{\rm def}{=} \left\{
\varepsilon^{\rm bulk}_n (k_x, k_y, k_z)|_{k_z}        \right\}
\end{equation}
The resulting distribution will show broad quasi-continuous bands and gaps 
(see Fig.~\ref{fig:projBS} in case of hexagonal graphite for 11 $k_z$ values). 

For bands in slabs in the limit of infinite thickness (but {\em without }
periodic macroscopic boundary conditions in $z$-direction)      
the following statements hold \cite{heine62}: \\
(i) Slab bands which lie in gaps of the PBBS are surface- or interface
bands  {\em localized} in $z$ direction toward the interior of the slab.\\
(ii) Slab bands which lie in (quasi-continuous) bands
of the PBBS agree exactly  in energy with the corresponding band of the  BBS,
and their density  agrees in the interior  
of the slab with  that of the bulk bands. \\
(iii) Slab bands within a gap, but close to the edge of a (quasi-continuous) 
band of the PBBS, have  weak localization (large decay length). 

Of course, numerical slab calculations are done on slabs of finite thickness. 
In case of any reasonable doubt concerning the character of the band,
 if  slab  bands are very close to band edges 
of the PBBS, one can obtain certainty only  by analyzing  
the spatial localization of the slab wave-function, e.g., 
by calculating the band weights.


\subsection{Total energy}
In order to check the possibility to use the LDA for total energy calculations 
of graphitic systems we first calculated the 
optimized lattice constants of bulk (AB) and (ABC) structures
and compared their total energies.
Despite the missing van-der-Waals correction, we observe in 
Table~\ref{tab-bulk-lattice-const } still some 
overbinding characteristic of the LDA, but all in all, the LDA does well
for the lattice constants.
From Table~\ref{tab-bulk-Etot } we can learn that the total energies
of hexagonal (AB) and rhombohedral (ABC) graphite are very
close in energy, whereas
hypothetical hexagonal (A) stacking is well above. The latter result
justifies the neglect of
stacking orders with two identical neighboring  layers  for
realistic sytems. 
However, (ABC) has a lower energy than (AB) despite the fact, 
that natural graphite is predominantly (AB) ordered.
Because the energy difference is less than one meV, we 
not only did these total energy calculations with an increased 
numerical precision of the overlap integrals~\cite{note1} 
and checked the convergence in the number of $\bf k$ 
points carefully (see below), but we tried also 
the GGA, the LDA with Perdew-Zunger XC, and we added an extra shell 
of orbitals to the default  basis set. None of these modifications 
reversed the ordering of the energies or changed essentially the energy
difference. 
Consequently, either the correct energetic order of
(AB) and (ABC) graphite 
is beyond the possibilities of the LDA and the GGA, 
or, the {\em electronic part} of the total energy in (ABC) is really 
lower than in (AB), 
but the here disregarded {\em phononic part} plays a decisive role 
for the ground state. 
In principle, there is also the possibility that 
the larger probability for (AB) ordering in natural graphite 
is due to special crystal growing conditions 
(temperature, pressure, etc.) and  that (AB) graphite 
is a meta-stable state under ambient conditions like diamond.

\begin{figure}[h]
\includegraphics[width=0.5\textwidth]{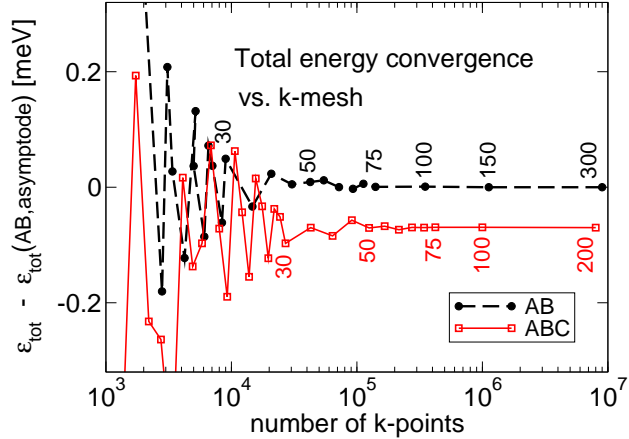}
\caption{ (color online)
Convergence of the total energy in LDA of (AB) and (ABC) graphite versus 
the total number of {\bf k} points in the BZ on a logarithmic scale. 
For both structures the optimized 
lattice constants as given in Table \ref{tab-bulk-lattice-const } 
were used. The reference energy is the asymptotic value for (AB) 
graphite.  The numbers next to some points for (ABC) graphite 
 indicate the number $N_1=N_2=N_3$ of k-points in each direction of the rhombohedral BZ. 
For hexagonal (AB) graphite the number $N_1=N_2$ of k-points in the plane parallel 
to the layers is given, whereby the number $N_3$ of k-points perpendicular to 
the layers has been chosen the closest integer of 1/3 of this number. 
The step width for $N_1$ is 1 for $N_1 \le 30$ and 5 for $30 \le N_1 \le 75$.
}
\label{fig:convergency} 
\end{figure}

The authors of
Ref.~\cite{gonze94} obtained for (AB) a total energy, 
which lies 0.1 meV {\em below} the total energy of (ABC) graphite.
They however used the Monkhorst-Pack procedure with 
28 special points (in the irreducible BZ) amounting to some hundred 
points in the full BZ. Fig.~\ref{fig:convergency} shows, that up to 
some 10 000 (equidistant) points for the tetrahedron method 
the total energies for both structures oscillate wildly 
making a save calculation of such a small energy difference impossible.
It is clear that the Monkhorst-Pack procedure has another
convergence behavior than the tetrahedron method, 
but on the other hand 
it is not clear at all, whether
it is suited for high precision demands on 
semi-metals like (AB) graphite, 
because it is tailored for semi-conductors and insulators.

\begin{table}
\begin{tabular}{|l|l|l|l|}
\hline
system & & exp.   & opt. \\ 
\hline
(AB)  & a      & 2.4595  & 2.450 \\
 & $c_{nn}$ & 3.33    & 3.306 \\
\hline
(ABC) & a      &   &   2.450 \\
 & $c_{nn}$ &   &      3.305 \\
\hline
\end{tabular}
\vspace{.5cm}
\caption{
Experimental and LDA-optimized lattice constants 
of bulk (AB) and (ABC) graphite in \AA.
The in-plane lattice constant is denoted $a$ and 
$c_{nn}$ is the distance between two layers.
The {\bf k} mesh for the optimization comprised 50 points in each dimension 
for both structures.
}
\label{tab-bulk-lattice-const }
\end{table}

\begin{table}
\begin{tabular}{|c|r|r|}
\hline
system & \; exp. \; &\; opt. \;   \\
\hline
(AB)  & 0 \; & -1.1  \; \\
\hline
(ABC) & -0.2 \; & -1.2    \;   \\
\hline
(A)   & +15.7 \; &     \\
\hline
\end{tabular}
\vspace{.5cm}
\caption{
Total energies per atom in meV for the experimental and the optimized
lattice constants (from Table~\ref{tab-bulk-lattice-const }) 
for bulk (AB), (ABC) and (A) stackings in LDA.
The reference energy is the energy for (AB) graphite  from the experimental  
lattice constant.
}
\label{tab-bulk-Etot }
\end{table}

\subsection{Estimate of the van-der-Waals correction}
In order to find out, if the consideration of the van-der-Waals interaction 
can clarify the situation, we estimated its impact 
on the total energies using the semi-empirical approach 
by Grimme {\it et al.}~\cite{grimme11} in the version DFT-D2 \cite{grimme06}: 
\begin{equation}
 E_D = s_6 \sum_R f(R) \frac{C_6}{R^6} 
\end{equation}
with the damping function 
\begin{equation}
 f(R) = [ 1 + e^{-d (R/R_0 -1) } ]^{-1}
\end{equation}
The sum runs over inter-atomic distances $R$ (calculated from experimental lattice constants of (AB) graphite) , 
the scaling factor is $s_6=1.1$, $d=20$, 
and for carbon the dispersion coefficient and the 
van-der-Waals radius amount to 
$C_6 = 1.75 \;\mbox{J nm}^6 \mbox{mol}^{-1} = 18.1\; \mbox{eV}\; \mbox{\AA}^6 $
and $R_0 = 1.452 \;\mbox{\AA}$, 
respectively.

We took into account only inter-layer contributions, because 
 the intra-layer contributions are independent of the stacking sequence.
It turns out, that the Grimme correction per atom 
to the total energy of bulk (AB) and (ABC) amounts  
to -210.4284 and -210.4289 meV,
respectively, providing a difference of 0.5 $\mu$eV in favor of 
(ABC) graphite.  
One reason for this minute difference
is the fact that the interaction energy of
two layers is independent of their type 
(provided they are not identical) and therefore only next-nearest 
(and beyond)  layer interactions
contribute to the energy difference between two stacking orders. 
Second, even if there is a difference, it is minute. 
There are only two values for the interaction between two arbitrary
layers depending on 
whether they are identical (say A-A) or different (say A-B). 
The interaction energy of the two atoms in layer A with a full layer A 
for next nearest neighbor layers 
is 12.161923 meV versus 12.161918 meV for A-B.
Consequently, the van-der-Waals correction at least
in the semi-empirical Grimme form can be safely 
neglected for the energy difference between stacking orders.


\section{Surface of graphite}

 Graphitic slabs are numerically demanding in three ways:  \\
(i) Because  graphite is a semi-metal with a tiny Fermi surface around 
the K-point, one needs a large number of $\bf k$ points in order to get the
essential features of the FS well resolved and an accurate position
of the Fermi level.
For the self-consistent calculation we used a $\bf k$ grid of 
240x240x1 inequivalent points.
For the LDOS near $\varepsilon_F$ we chose a special grid of points,
which is restricted to the vicinity of the K-point and comprises
150x150x1 points.\\
(ii) Due to the small density of states near $\varepsilon_F$
the screening of perturbations may extend over long distances.
Thus, the slabs used in the numerical calculation have to be choosen
rather thick, if separated interfaces or surfaces shall be described.
For slabs, which are thick enough, extra  layers within the given building
scheme  should not have an impact on the physical results. \\
(iii)  Thick slabs with low symmetry tend to have  problems
in converging to self-consistency. Therefore, it is vital to use the 
highest possible symmetry.  
This can be achieved 
by choosing an appropriate number of layers (see Sect.~IV-VI). 

\subsection{Band structure of (AB) slabs}

\begin{figure}[h]
\centering
\includegraphics[width=0.5\textwidth]{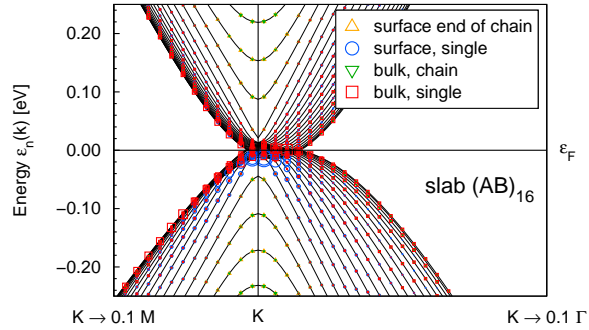}
\vspace{-1cm}
\caption{(color online) Band weights of the local orbital $2 p_z$ 
located at the surface and in the middle (denoted 'bulk') of the 
slab (AB)$_{16}$.
}
\label{fig:BW-interface-bulk-AB_16}
\end{figure}

\begin{figure}[h]
\includegraphics[width=0.5\textwidth]{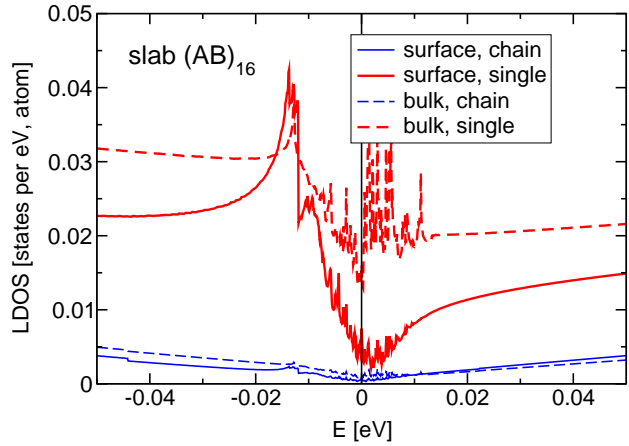}
\caption{(color online) 
Surface versus bulk LDOS (sum of all orbitals per lattice site)
in the $(AB)_{16}$ slab.  }
\label{fig:slab-LDOS-AB_16}
\end{figure}

We consider a  clean surface, 
which  can be studied by means of a thick slab.
Fig.~\ref{fig:BW-interface-bulk-AB_16} shows band weights in (AB)$_{16}$ 
for a few prominent local orbitals  and Fig.~\ref{fig:slab-LDOS-AB_16} 
presents the corresponding LDOS. 
For a visualization of the atomic positions in the slab see the right
part of Fig.~\ref{fig:atoms-ABC}. As to $p_z$-bonding, 
which is crucial for the electronic structure around the Fermi energy,
we again distinguish single atoms (monomers) and  chains, which are now finite.

First, in Fig.~\ref{fig:BW-interface-bulk-AB_16} we observe no 
low-energy surface bands separated from the bulk continuum. 
This  refers to  higher energies in the range of  the $\pi$ and $\sigma$ bands
on the symmetry lines as well (not shown in our figures) 
in accordance with Ref.~\cite{ooi06}. 
In contrast, the LDOS in Fig.~\ref{fig:slab-LDOS-AB_16} reveals that
around $\varepsilon_F$ there is  a {\em depletion} of 
electrons in the surface layer rather than
an accumulation (which would be expected for surface states).
It should be expected that the bulk LDOS 
in Fig.~\ref{fig:slab-LDOS-AB_16} converges to the 
bulk curves shown in Fig.~\ref{fig:bulk-LDOS-AB}  
in the limit of infinite thickness of the slab. 
This is already suggested by Fig.~\ref{fig:slab-LDOS-AB_16},
except for the peak just above the Fermi energy $\varepsilon_F=0$,
which is decomposed into a bunch of single peaks.
Closer inspection of the band structure 
in the energy region around  $\varepsilon_F$ (not shown)  reveals that 
these peaks are mainly due to van-Hove singularities caused by avoided 
crossing of bands, which gradually 
disappear if the number of bands goes to infinity. \\
Second, Fig.~\ref{fig:slab-LDOS-AB_16} also shows that 
the bands near $\varepsilon_F$ are mainly localized at the single atoms. 
Considering the results of the previous paragraph, 
the latter conclusion is not surprising because the band structure 
of a thick slab must be similar to the PBBS, if no 
surface bands exist. Fig.~\ref{fig:slab-LDOS-AB_16} shows that 
the LDOS around $\varepsilon_F$ at single atoms in the bulk 
is one order of magnitude larger than at chain atoms 
in agreement with the bulk calculation in Fig.~\ref{fig:bulk-LDOS-AB}.

In terms of the local conductivity these two points  lead to the conclusion 
that one should expect a depletion of conductivity in the surface layer and 
current flow mainly through hopping between single atoms. 
The strong asymmetry regarding the two atomic positions at the surface 
 in the LDOS near the Fermi energy explains also the strong asymmetry 
 of the two sites  in STM images \cite{tomanek87}.

\subsection{Surface energy}

\begin{figure}[h]
\includegraphics[width=0.5\textwidth]{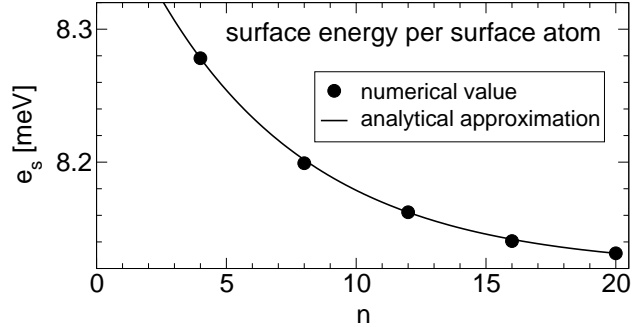}
\caption{
Convergence of the surface energy per surface atom versus the number of unit cells 
in slabs (AB)$_n$. 
}
\label{fig:surface-energy} 
\end{figure}

We use the standard definition of the surface energy per surface atom 
\begin{equation}
 e_s = \frac{1}{4} (E_{sc} - N_{at} \cdot e_b) 
\label{surface-energy}
\end{equation}
where $E_{sc}$ is the total energy of the super-cell with $N_{at} $ atoms 
and $e_b$ is the total energy per atom for the bulk. Because each super-cell has
two surfaces and two surface atoms on either surface,
we introduced the factor $1/4$ in the definition.
In principle the result of Eq. (\ref{surface-energy}) might depend on 
the thickness of the layer, but has to converge 
in the limit of infinite thickness, provided, $E_{sc}$ is 
calculated in the interior of the slab 
with the same precision as $e_b$ \cite{boettger94}. 
We met this demand by using the same program with the same parameters 
and the same $\bf k$ grid in the plane parallel to the layers 
(120x120x50 and 120x120x1 for the bulk and the slabs, respectively). 
Our results shown in Fig.~\ref{fig:surface-energy} are not yet fully converged, but 
a tendency toward convergence is obvious.
In order to obtain an approximate asymptotic value $ e_\infty$, 
we adapted the three parameters of the ansatz 
$e_s(n)= e_\infty + a  \; exp(-b\;n)$ to the three calculated values at 
$n$ = 4, 12, and 20, where our slab (AB)$_n$ consists of $n$ formula units.
We found 
$ e_{\infty} = 8.12$ meV, $a = 0.305$ meV, and $b=0.165$. 
The corresponding curve is plotted in Fig.~\ref{fig:surface-energy}.
Experimental values for (AB) graphite lie  in the wide range of
(2.83 - 18.9) \mbox{meV/surface atom} (see Table 5 in \cite{ooi06})
and the result for $n= 12$ using Eq.~(\ref{surface-energy}) and the VASP code is
7.92 \mbox{meV/surface atom}~\cite{ooi06}.

Although we cannot reach full numerical convergence of $e_s$
with layer thickness, it is worthwhile 
to compare surface energies for the surfaces of (AB) and (ABC) graphite 
for the same number of layers. 
Calculations for  slabs with super-cells  
(AB)$_6$ and (ABC)$_4$ (both have 12 layers) provide 
$e_s = 8.23$ meV and $e_s = 8.38$ meV, respectively. 
From a local point of view the smallness of the 
difference between the two systems seems understandable,
because  both surfaces  
differ only in the third layer from the surface inward. 
On the other hand, both systems differ qualitatively in their
electronic structure. 
Whereas (ABC) graphite has around $\varepsilon_F$ 
topologically protected surface states and a
Dirac-like band structure in the bulk limit~\cite{xiao11},    
(AB) graphite is a semi-metal with a small Fermi surface and 
without surface states in the whole region of $\pi$ and 
$\sigma$ bands. 

\section{ Stacking faults}

For the calculation of one-particle energies and derived properties we used 
a periodic arrangement of  slabs which
are surrounded by vacuum and which have {\bf one}  stacking fault in the middle.
The alternative model is a periodic arrangement (without vacuum)
with {\bf two} stacking faults per unit cell in order to obtain periodicity,
which has been adopted for total energies (see Introduction).  
We want to stress that for the size of slabs  presented here there is no 
remarkable difference between both models in the 
local properties like LDOS and band weights 
for atoms close to the stacking faults.
All self-consistent  calculations on slabs have been done with a 
$\bf k$ mesh of 120x120x1 points equal-distantly distributed over the BZ.
Test calculations with 240x240x1 points did not show any visible changes.
For the calculations on periodically repeated super-cells 
we used  a grid of 150x150x10 points.

\begin{figure}[h]
\includegraphics[width=0.5\textwidth]{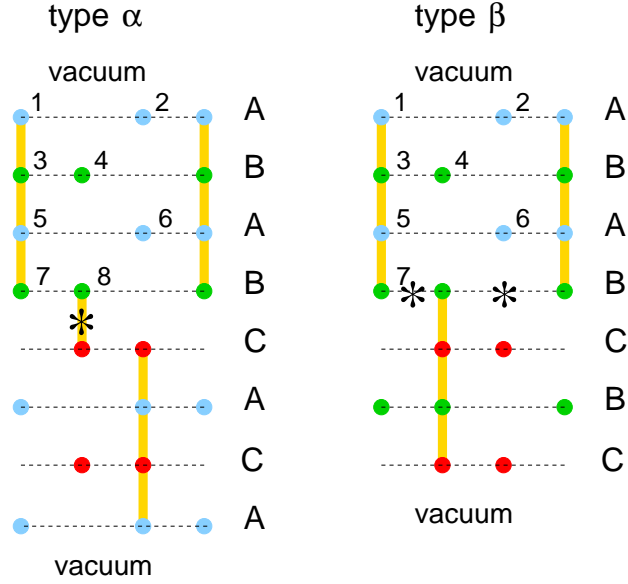}
\caption{(color online) 
Examples for the two types of stacking faults in highly symmetric slabs 
separated by vacuum slabs. 
Left panel: type $\alpha$: (AB)$_n$(CA)$_n$ and
right panel: type $\beta$: (AB)$_n$(CB)$_{n-1}$C, both for $n=2$.
(The numerical calculations presented below were done for $n=8$.)
The yellow perpendicular bars indicate saturated $p_z$ bonds and
the stars mark inversion centers.
Both types of slabs have the symmetry $P\bar{3}m1$.}
\label{fig:structure-slabs}  
\end{figure}

Fig.~\ref{fig:structure-slabs} gives the essential structural information 
on the two possible types of stacking faults in hexagonal graphite,
denoted $\alpha$ and $\beta$.
The number of layers has been chosen
in such a way that the highest possible symmetry (and therefore precision
for given computational resources) could be achieved.
Whereas bulk hexagonal (AB) graphite has two Wyckoff positions, 
namely single atoms (monomers) 
and atoms on infinite chains with saturated $p_z$ orbitals,  
the chains are terminated at the surface and the interfaces 
ending with dangling bonds. The crucial difference
as to local bonding  between type $\alpha$ and $\beta$ interfaces
is the following:  
Type $\alpha$ has a dimer
bridging the stacking fault, with a related small interlayer overlap
$\gamma_1$ and indirect overlap between the chains on both sides of the
fault via the dimer. In type $\beta$, the dangling bonds of
two adjacent chains overlap laterally with the large overlap integral
$\gamma_0$.

\subsection{Band structure}

\begin{figure}[h]
\centering
\includegraphics[width=0.5\textwidth]{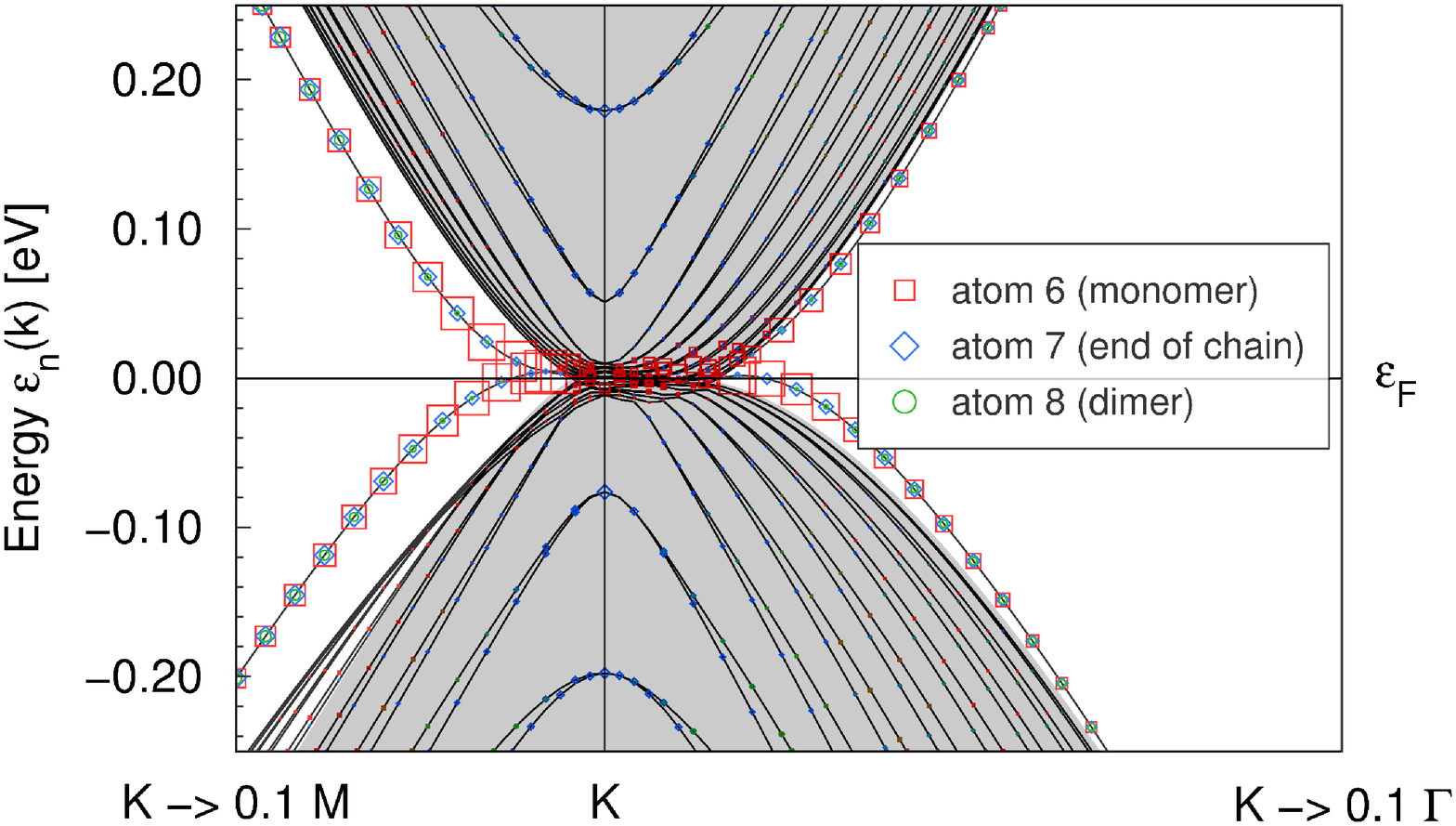}
\vspace{-.5cm}

\includegraphics[width=0.5\textwidth]{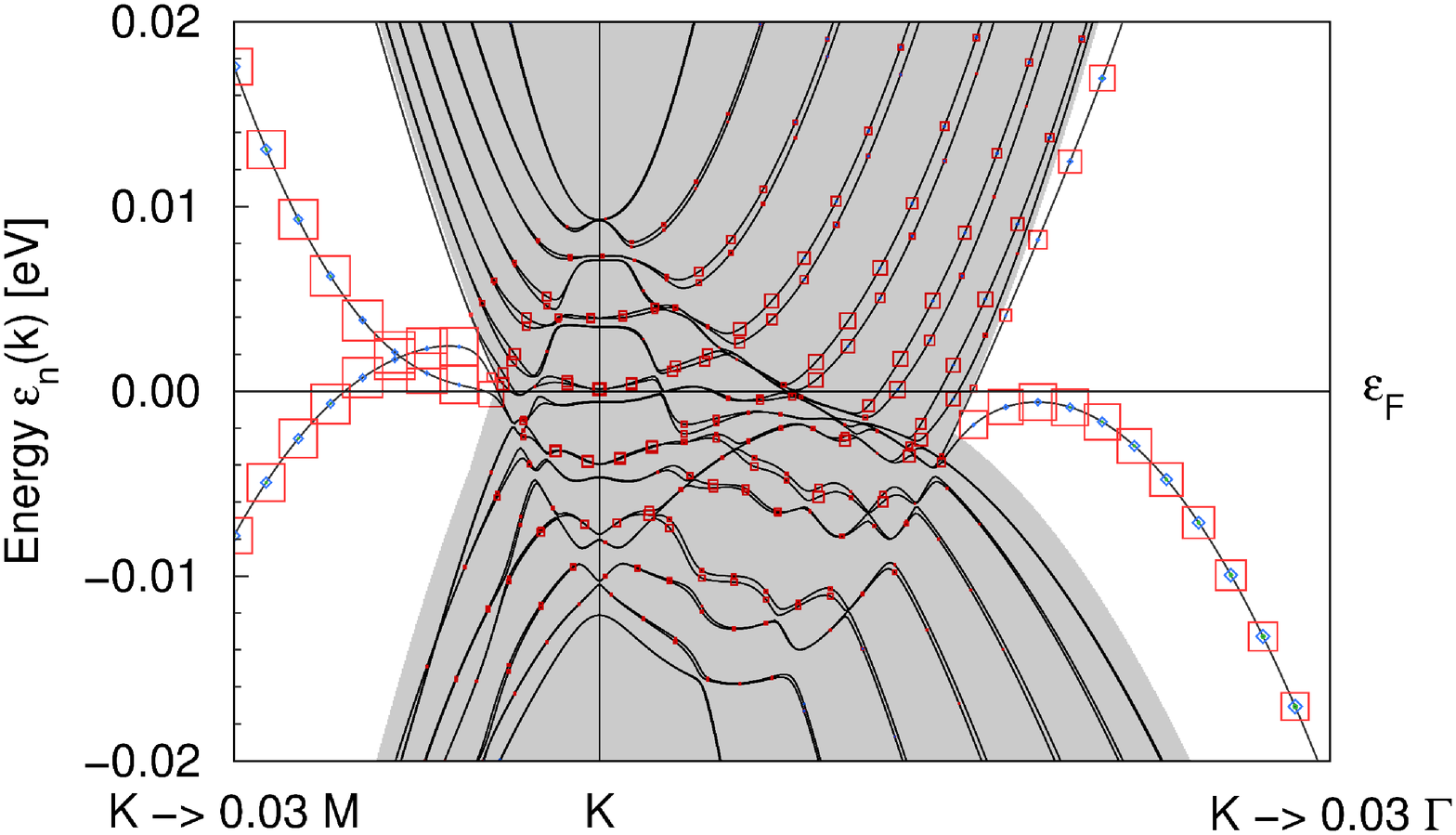}
\caption{(color online)
Band weights of the orbital $2 p_z$ 
close to a stacking fault of type $\alpha$  for  (AB)$_8$(CA)$_8$ 
with two energy scales.
The atom numbers 6 to 8 refer to atoms in the two layers next to the fault,
as defined in Fig.~\ref{fig:structure-slabs}. }
\label{fig:BW-interface-alpha}    
\end{figure}

\begin{figure}[h]
\centering
\includegraphics[width=0.5\textwidth]{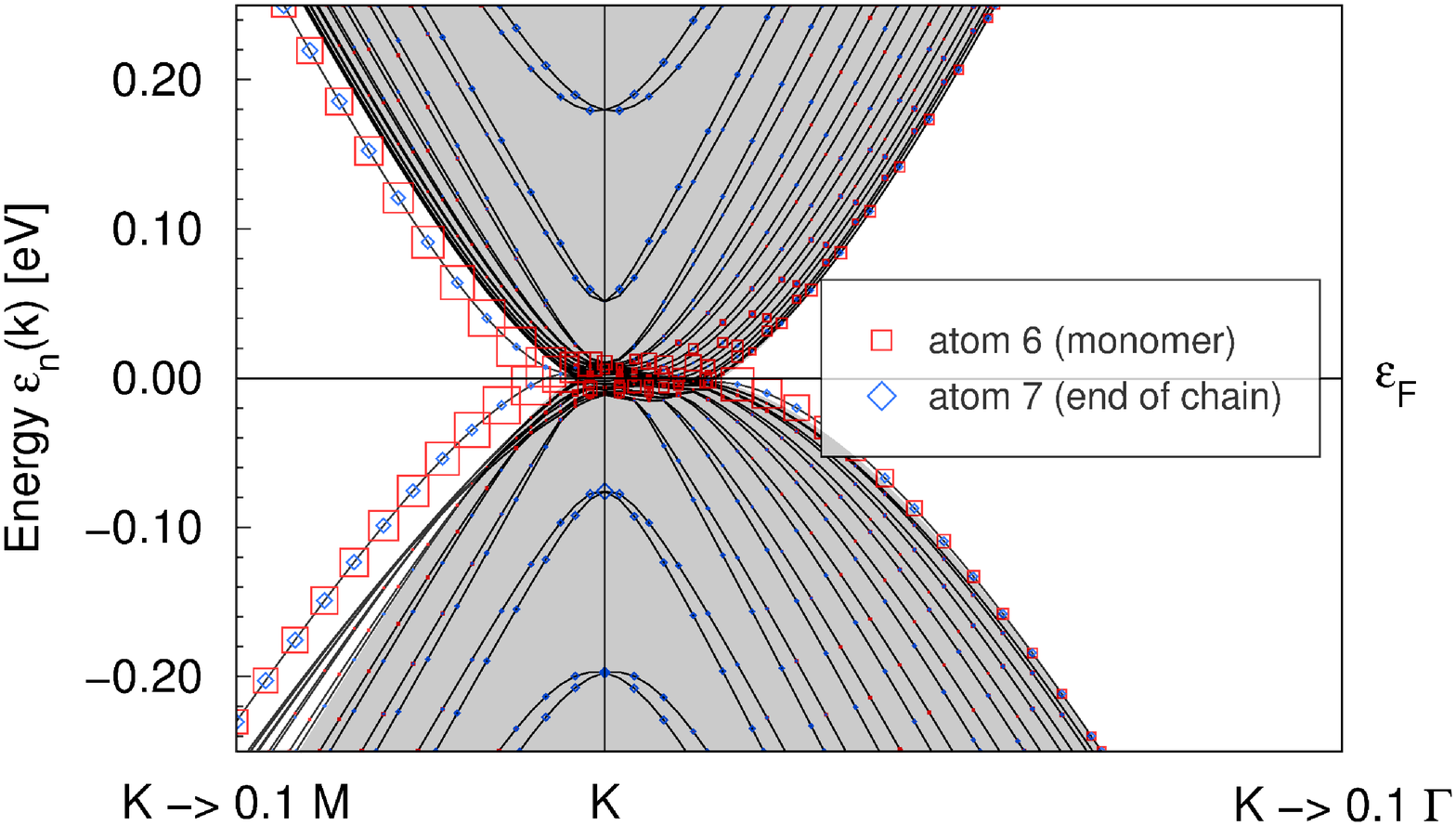}
\vspace{-.5cm}

\includegraphics[width=0.5\textwidth]{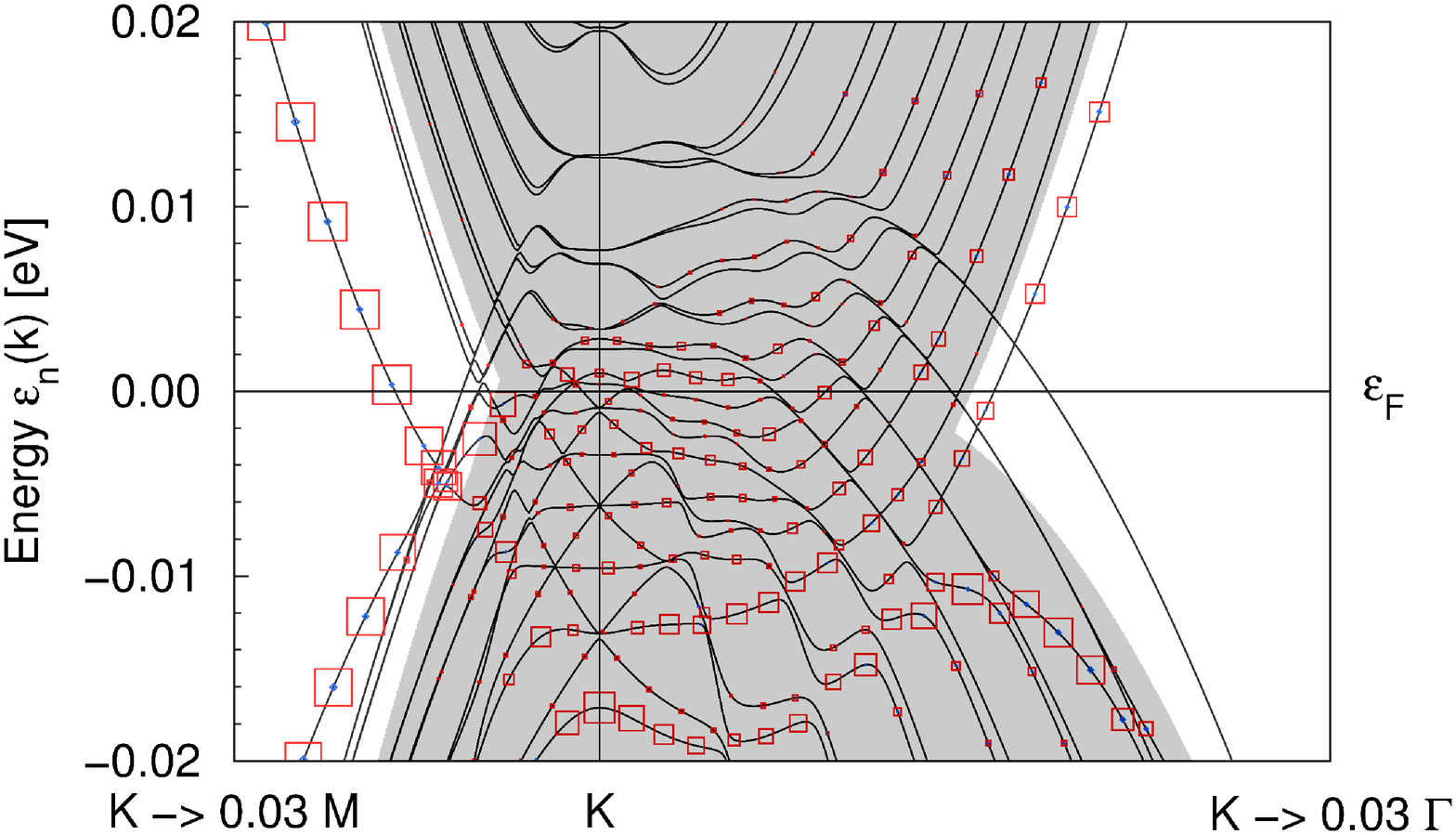}
\caption{(color online) 
Band weights of the orbital $2 p_z$ close to a stacking fault of type 
$\beta$ for  (AB)$_8$(CB)$_7$C.  
Further details as in Fig.~\ref{fig:BW-interface-alpha}.
}
\label{fig:BW-interface-beta}    
\end{figure}

Figs.~\ref{fig:BW-interface-alpha} and \ref{fig:BW-interface-beta}
show the energy bands with the most prominent band weights 
for $\alpha$ and $\beta$ interfaces around the K-point. 
We observe that for both interfaces there is an occupied
and an empty interface band 
close to the bulk continuum, except in the very vicinity 
of the K point. In either case, the interface bands are located 
mostly at the monomers closest to the interface 
(atom \# 6 in Fig.~\ref{fig:structure-slabs}), i.e., they are formed by
the dangling $p_z$-bonds of these monomers. 
The band weights of the  monomers further away from the interface 
(e.g. \# 4 in Fig.~\ref{fig:structure-slabs}) in the interface bands 
are already small and would  not be visible in these figures.
This means that the amplitude of the interface bands converges rapidly to zero 
toward the bulk. 
Exceptions are the interface bands on the line K-$\Gamma$ in type  $\beta$. 
Their distance from the bulk continuum is smaller 
and their localisation is weaker than for the other interface bands. 
Therefore, the existence of an interface state in this $\bf k$ region 
in the limit of infinite thickness is not absolutely certain.

Comparison of our results for type $\beta$ with the model calculations in 
Refs.~\cite{guinea08} and \cite{koshino13}
shows little similarity apart from the mere existence of interface states.
The main difference is that in our calculation the interface states 
vanish in the very vicinity of the K-point 
due to  the presence of bulk states. 
The low-energy dispersion of these bulk states is not correctly described 
by simple models with only two overlap parameters.

\begin{figure}[h]
\centering
\includegraphics[width=0.5\textwidth]{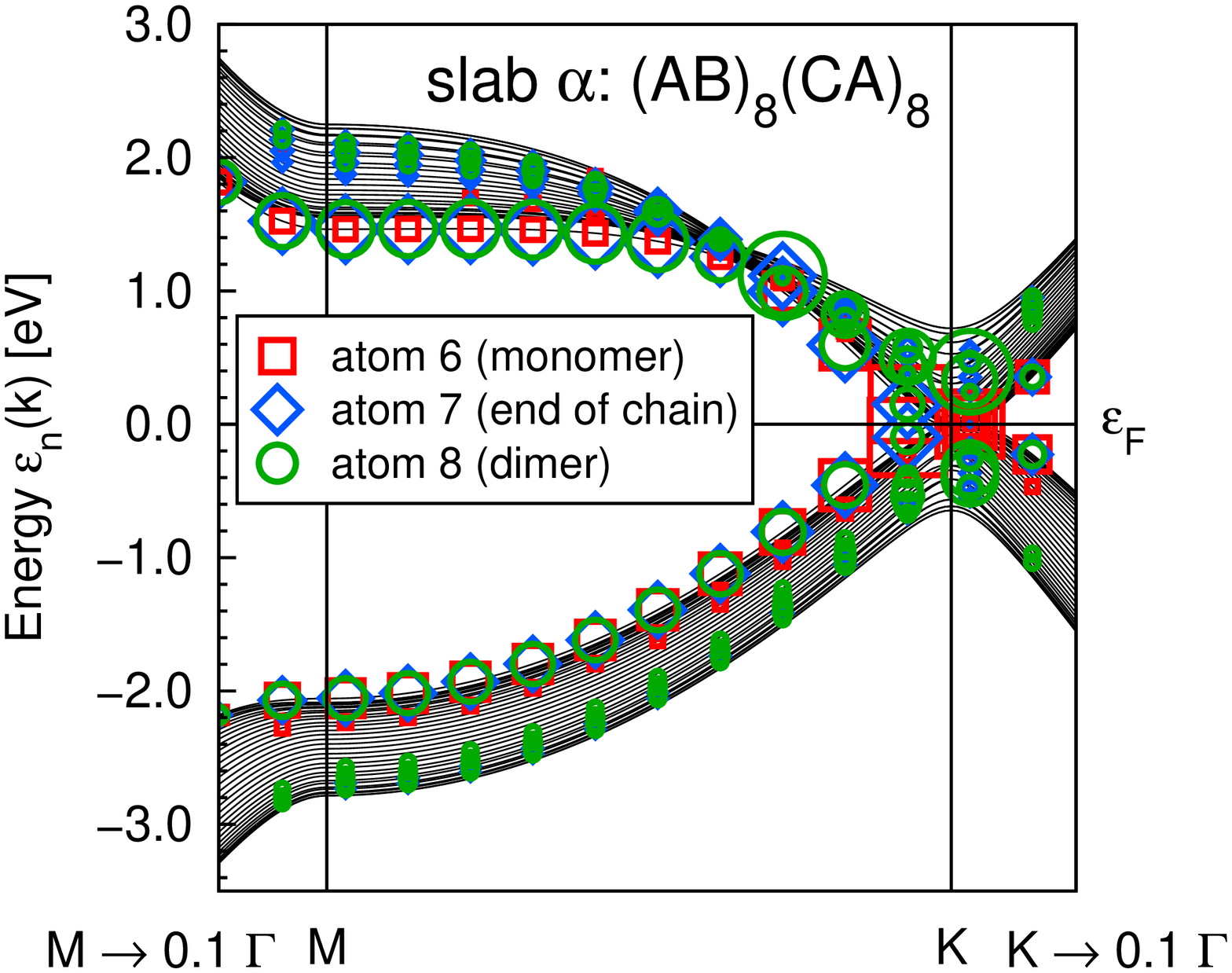}
\includegraphics[width=0.5\textwidth]{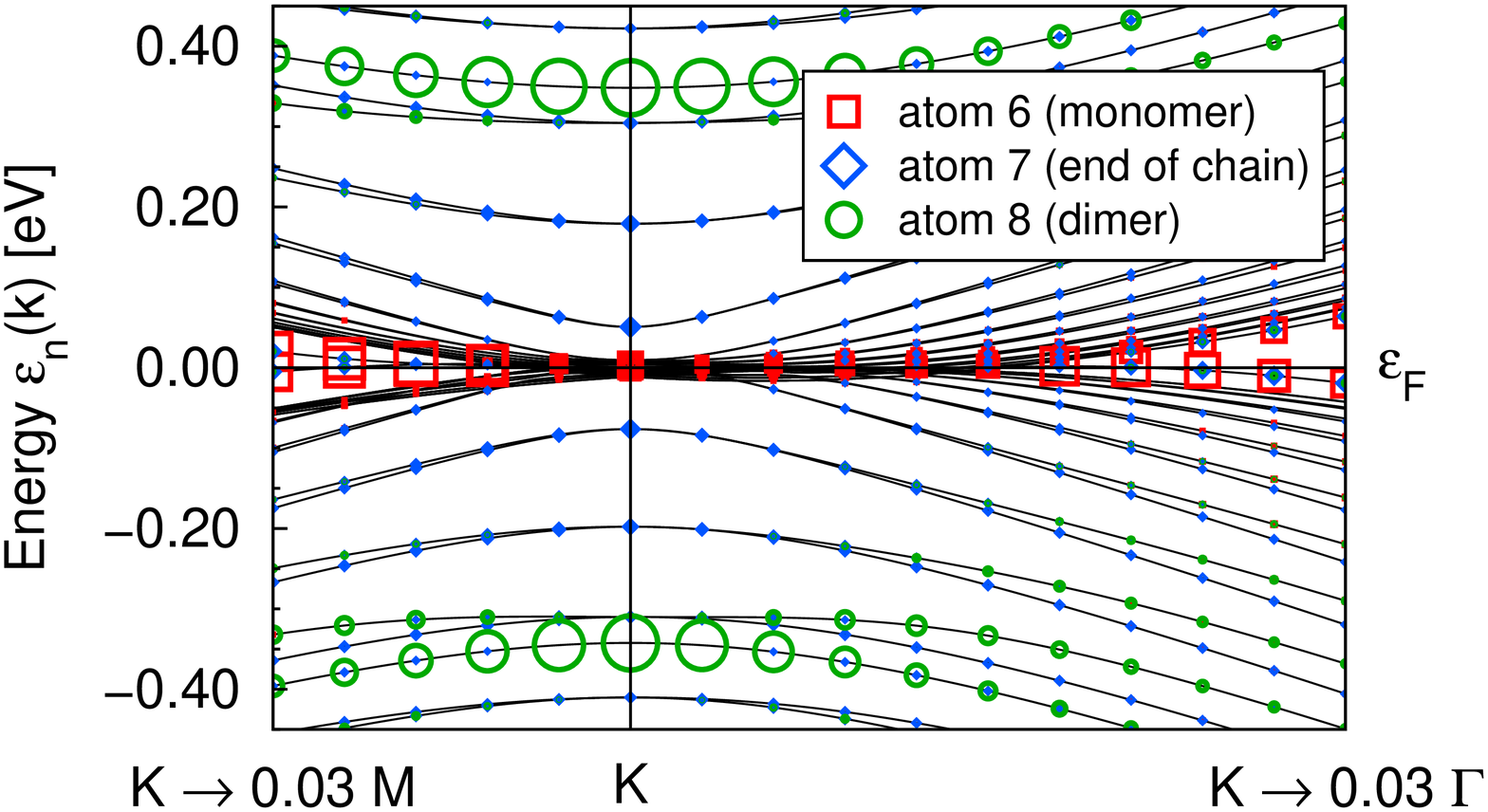}
\caption{(color online) Band weights of the orbital $2 p_z$ 
close to a stacking fault 
of type $\alpha$ for  (AB)$_8$(CA)$_8$. 
Upper panel: line K-M and parts of the adjacent symmetry lines,
lower panel: blow up 
of the lines K-M and K-$\Gamma$ around the point K. 
For a better overview, in the upper panel all symbols are omitted, 
which are smaller than 50\% of the maximum size.  
The big green circle in the conduction bands just right of the vertex of bands 
is not a numerical fluctuation, but an indication for a
genuine strong variation of the band weights in this region 
of the $\bf k$ space. 
}
\label{fig:BW-dimers-AB_8CA_8}
\end{figure}

Fig.~\ref{fig:BW-dimers-AB_8CA_8} shows another interesting feature. 
There are occupied and empty  bands localized at the dimers 
of interface $\alpha$, which are split by approximately 0.8 eV at the K-point 
(see lower panel).  
This splitting agrees with the value,
which is expected for the splitting of the $p_z$ levels 
in an {\em isolated dimer} $ E_\pm = E_0 \pm \sqrt{|\gamma_1|^2}$ due to 
an empirical overlap integral 
with the generally accepted value $\gamma_1 \approx 0.4$ eV.
Near the K-point,
these dimer bands are no strictly localized interface bands,
but resonances with a large amplitude 
near the interface,
because they are submerged into the  the continuum of bulk states. 
The upper panel shows, that  along the line from K toward M  
 not only  the dimer atoms get involved in the interface bands, but 
also the atoms at the end of the chains.
Therefore, away from the K-point 
and at energies of the eV-range, the interface bands are no longer localized
solely at monomers. 
Inspection of the other parts of the 2D BZ (which are not shown in the figures) 
shows that {\em on the symmetry lines} interface states  
are only found on  the line K-M and 
on small parts of the lines M-$\Gamma$ and K-$\Gamma$. 
In other words, they are virtually 
restricted  to the symmetry lines shown in the upper panel of
Fig.~\ref{fig:BW-dimers-AB_8CA_8}.

\begin{figure}[h]
\centering
\includegraphics[width=0.5\textwidth]{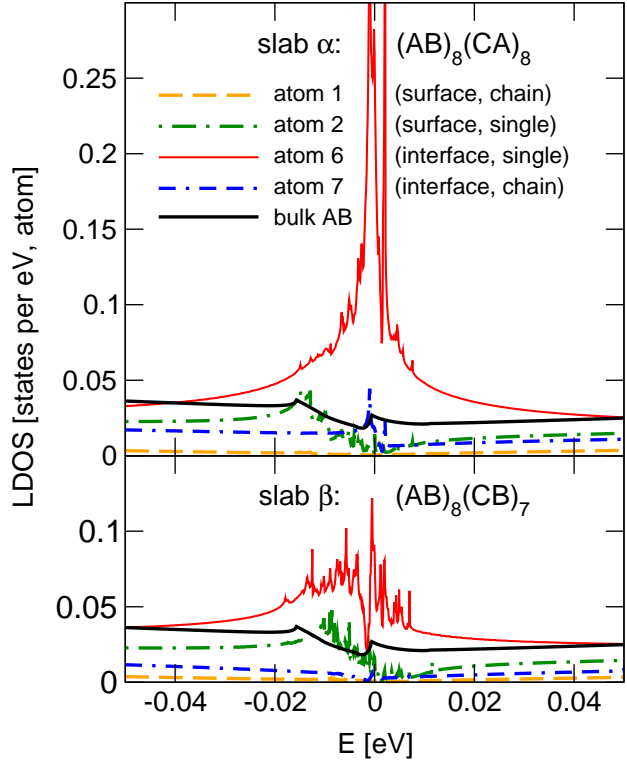}
\caption{(color online) LDOS (sum of all orbitals per site) at atoms near 
the surface and near the interface   
in a slab of type  $\alpha$  for (AB)$_8$(CA)$_8$ (upper panel) 
and a slab of type  $\beta$ for (AB)$_8$(CB)$_7$C (lower panel) 
compared with the bulk DOS. 
For the LDOS a special mesh 
of 150x150x1 points located in a restricted area around the K point was used.
}
\label{fig:LDOS-interface-surface}
\end{figure}

Fig.~\ref{fig:LDOS-interface-surface} 
presents the LDOS 
near both types of interfaces and near the surfaces compared with the bulk DOS.
In either case the interface bands produce strong low-energy peaks 
in the LDOS for the monomers closest to the interface,
but in type $\alpha$ the peak is most pronounced. 

\begin{figure}[h]
\centering
\includegraphics[width=0.5\textwidth]{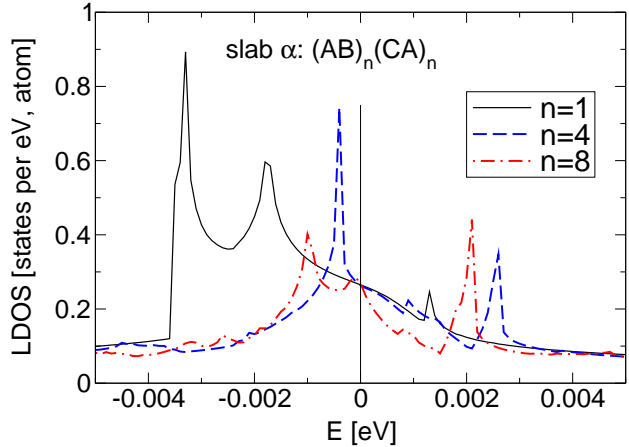}
\caption{(color online) LDOS (sum of all orbitals per site) at the atoms \# 6 
(monomer closest to the stacking fault) 
in slabs (AB)$_n$(CA)$_n$ for several values of $n$.
(Observe the zoomed scale of the E-axis compared with the other LDOS-figures.)
}
\label{fig:LDOS-interface-surface-AB_nCA_n}
\end{figure}

Fig.~\ref{fig:LDOS-interface-surface-AB_nCA_n} presents the LDOS 
of slabs with interface $\alpha$ for different thickness of the slab.
One notes that even a slab with one graphite unit cell on both sides of 
the stacking fault (n=1) shows a pronounced peak and is thus a suited model for studying 
the interface bands, although the specifics 
of the low-energy electronic structure depend on the slab thickness, 
which is seen in the form of the LDOS. 

\subsection{Formation energy of stacking faults}

\begin{figure}[h]
\includegraphics[width=0.5\textwidth]{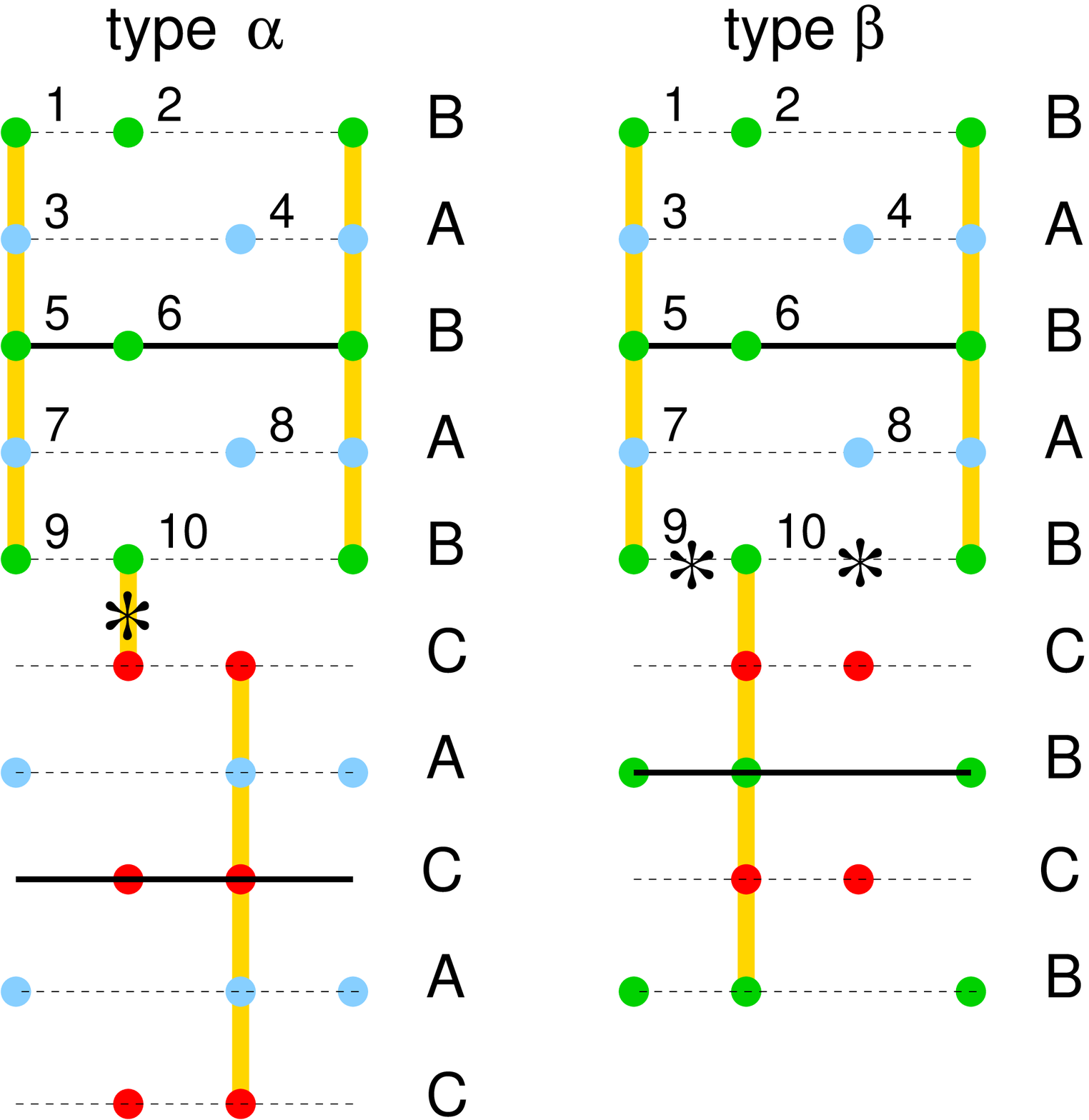}
\caption{Examples for the two types of stacking faults in highly symmetric
super-cell geometry without surfaces.
Type $\alpha$: B(AB)$_n$(CA)$_n$C and
type $\beta$: (BA)$_n$(BC)$_n$, each for $n$=2.
(The numerical calculations presented here have been  done for $n$ = 4 and 8.)
The stars denote inversion centers and the full lines mirror planes.
Observe that the lowest layer in the right panel does not contribute to the
formula for the unit cell because it is at the bottom of 
the cell and therefore equivalent to the highest, 
whereas all layers of the left panel 
are in the interiour of the cell.
}
\label{fig:structure-supercells}  
\end{figure}

The total energies of systems with stacking faults were calculated using super-cells 
without surfaces (see Introduction). For stacking fault $\alpha$  and $\beta$ the 
unit cells  B(AB)$_n$(CA)$_n$C and (BA)$_n$(BC)$_n$, respectively, 
with $n$ = 4 and 8 were used (see Fig.~\ref{fig:structure-supercells}).
They  differ slightly from the  cell in the geometry with surfaces, 
because this allows to retain 
the high symmetry of bulk AB graphite (group $P6_3/mmc$).
The difference concerns only the exact number of layers per block, which
should be irrelevant for the electronic structure around the stacking faults, 
if the blocks are thick enough.   

The interface contribution to the total energy per interface atom is defined 
 in analogy to the surface energy Eq.~(\ref{surface-energy})  as 
\begin{equation}
 e_{if} = \frac{1}{4} ( E_{sc} - N_{at}\cdot  e_b) 
\end{equation}
The factor $1/4$ comes from the fact that we have two stacking faults per unit 
cell and two atoms in the interface layer.
The interface energy $e_{if}$ shown in Table~\ref{tab1} shows  
that the total energy of a slab with interface $\alpha$
lies only slightly above pure bulk, 
whereas the extra energy of interface $\beta$ is somewhat larger.
The values also depend on whether
we use the experimental or the theoretical lattice constants.   
We further observe, that  in the range of 18 to 34 layers the 
interface energy still depends slightly on the width of the 
bulk blocks, which is in agreement with the convergence behavior
of the surface energy discussed in Sect.~III B. 

The energetic advantage of stacking fault $\alpha$ versus $\beta$ 
could be understood with the existence of the (occupied)
 dimer band in the former, 
which lowers the sum of one-particle energies as an important part of the total energy.
If  the electronic part of the total energy is a measure for realization of the structure, 
this result would make interface $\alpha$ more likely to occur 
in real crystals than interface $\beta$.
Consider, however, that the 
energy differences of the electronic part are of order $10^{-5}$ eV,
and that the effects discussed in Sect.~II C can have a strong impact as well.

\begin{table}
\begin{tabular}{|c|c|c|}
\hline
system & formula & $e_{if}$ \\
\hline
$\alpha$ & B(AB)$_4$(CA)$_4$C  &   9.5  \\
          &                  &  (23.4) \\
\hline

$\alpha$ &B(AB)$_8$(CA)$_8$C   &  7.9    \\
\hline

$\beta$  & (BA)$_4$(BC)$_4$    &    34.6 \\
          &                  &  (43.5) \\
\hline

$\beta$   & (BA)$_8$(BC)$_8$    &   33.5   \\
\hline

\end{tabular}
\vspace{1cm}
\caption{
Interface contribution to the total energy per interface atom, $e_{if}$, as
defined in the text (in $\mu$eV) for hexagonal graphitic stacks in LDA. 
For the upper line in each block, the experimental 
lattice constants of (AB) graphite are used for 
all stacking sequences, and for the numbers in parentheses 
the optimized theoretical lattice constants
from Table~\ref{tab-bulk-lattice-const } are applied. 
For the matrix elements the increased precision \cite{note1} is applied throughout.
}
\label{tab1}
\end{table}


\section{Displaced surface layer }

\begin{figure}[h]
\includegraphics[width=0.25\textwidth]{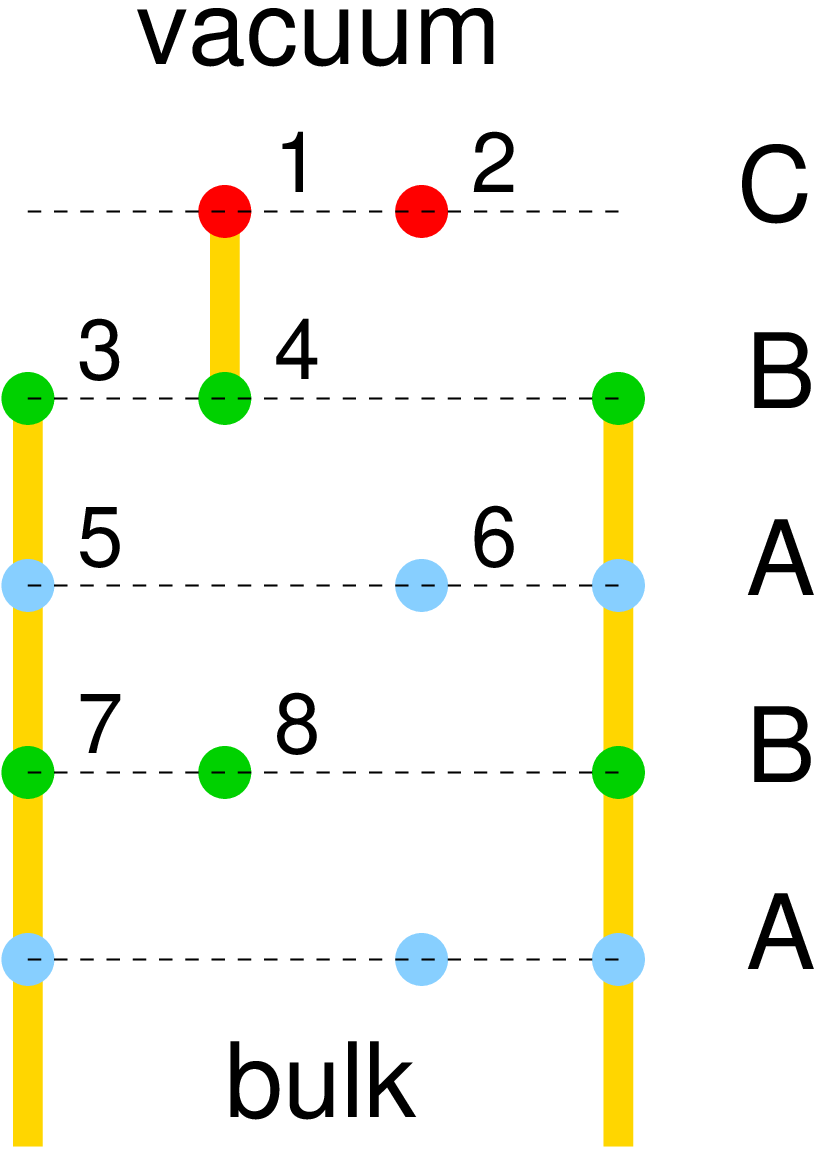}
\caption{(color online) Surface region of a slab (CB)(AB)$_{n}$(AC) 
modeling hexagonal graphite with a displaced surface layer on either side.
}
\label{fig:structure-overlayer}  
\end{figure}

\begin{figure}[h]
\centering
\includegraphics[width=0.5\textwidth]{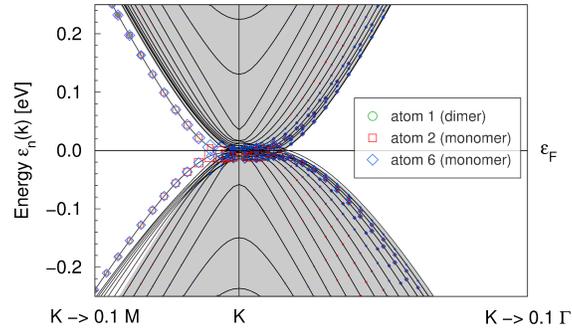}
\caption{(color online) Band weights of the orbital $2 p_z$  near a 
displaced surface layer in 
the slab (CB)(AB)$_{10}$(AC).}
\label{fig:overlayer}    
\end{figure}

We have shown above that there are no surface bands in hexagonal graphite, but 
stacking faults can induce interface bands. The next issue is, if only {\em one}
displaced surface layer with the geometry shown in Fig.~\ref{fig:structure-overlayer}
can do this job.
The resulting band structure with band weights in Fig.~\ref{fig:overlayer} 
shows that at least 
on the line K - M there is a surface band which is mostly localized at the 
monomers 2 and 6 defined in Fig.~\ref{fig:structure-overlayer}. 
On the line K - $\Gamma$ there are surface resonances with increased weights
toward the surface, but no real surface bands.
Note that the structure has dimers at the surface which 
produce localized dimer bands (not shown) split by approximately 0.8 eV 
similar to the dimers near stacking faults. 
These dimer bands look very similar to those shown in the lower panel of 
Fig.~\ref{fig:BW-dimers-AB_8CA_8} for a slab with stacking fault $\alpha$,
but each dimer band is almost doubly degenerate with 
a slight splitting. This is due to the existence of two dimers in the 
unit cell of the slab 
(one on either surface)  with a small interaction through the slab.   

\begin{figure}[h]
\centering
\includegraphics[width=0.5\textwidth]{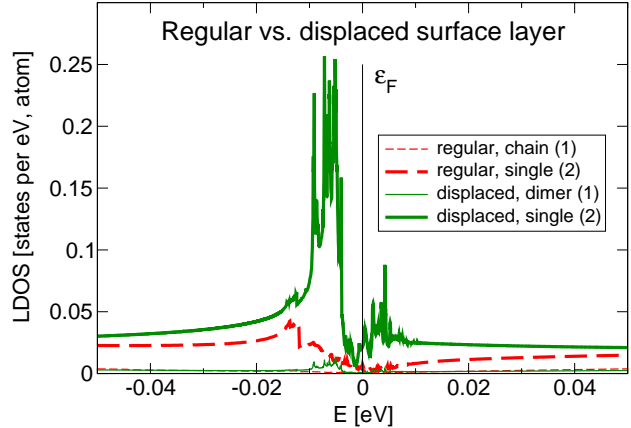}
\caption{(color online)
LDOS (sum over all orbitals per site) at atoms on a regular
and a displaced surface layer in slabs (AB)$_{16}$ and (CB)(AB)$_{10}$(AC), 
respectively. The numbers in the legend refer to the atom number in 
Figs.~\ref{fig:structure-slabs} and \ref{fig:structure-overlayer}.  
}
\label{fig:overlayer_DOS}    
\end{figure}


\section{Summary}

We investigated the electronic structure of hexagonal graphite without and with 
surfaces and stacking faults using self-consistent
full-potential DFT calculations in the LDA and the GGA.
There are two types of stacking faults (denoted by $\alpha$ and $\beta$)
which differ in the first place by their chemical bonding of the $2 p_z$ orbitals
in the vicinity of the stacking fault (see Fig.~\ref{fig:structure-slabs}).
We find that

\begin{itemize}
\item
{\em Pure surfaces  } do not host any surface bands in the energy range
 of the $\pi$ and $\sigma$ bands. 
Because the LDOS  around $\varepsilon_F$  in the surface layer is reduced, 
we expect  a reduced surface conductivity compared with the bulk.
Both, at the surface and in the bulk, the LDOS around 
$\varepsilon_F$ is one order of magnitude larger for the single atoms 
than for the chain atoms.
Therefore, 
all low-energy electronic
properties (like electric conductivity, low-temperature thermal conductivity
and specific heat) are governed by hopping processes between the dangling
$p_z$-orbitals of the single atoms.

\item Even displacement of {\em one} single atomic layer at the surface,
which is the germ for producing a stacking fault in the crystal growing process, 
can induce surface bands.

\item
Both types of {\em stacking faults} induce  interface bands
around the K-point in the Brillouin zone.
Correspondingly, the LDOS at the single atoms near a  stacking fault 
is enhanced over the bulk value. In the case of type $\alpha$  
the enhancement factor  is of the order of  10. 
This indicates a large 2D electronic conductivity along the stacking fault. 

\item
Stacking faults of type (AB)$_n$(CA)$_n$ (denoted by $\alpha$) 
are characterized by $p_z$-bonded dimers, 
which produce a pair of dimer bands. They are split by about 0.8 eV at the K-point 
and could be probed by near-infrared spectroscopy. 
Within the LDA, their electronic part of the  formation energy is 
smaller than for the alternative type (AB)$_n$(CB)$_{n-1}$C (denoted by $\beta$).

\end{itemize}

Our results indicate that it can be extremely misleading 
if experiments on real graphite samples 
(which have most likely numerous stacking faults) 
are compared with electronic structure calculations 
on ideal lattices.  This applies in particular to transport measurements which are governed by 
the electronic structure around the Fermi energy.

\begin{acknowledgments}
We are indebted to J. van den Brink, J. Venderbos, P. Esquinazi and 
M. Knupfer for helpful discussions.
Financial support was provided by DFG Grant RI932/6-1.
\end{acknowledgments}


\bibliographystyle{apsrev}  
\bibliography{ref.bib}

\begin{thebibliography}{34}
\expandafter\ifx\csname natexlab\endcsname\relax\def\natexlab#1{#1}\fi
\expandafter\ifx\csname bibnamefont\endcsname\relax
  \def\bibnamefont#1{#1}\fi
\expandafter\ifx\csname bibfnamefont\endcsname\relax
  \def\bibfnamefont#1{#1}\fi
\expandafter\ifx\csname citenamefont\endcsname\relax
  \def\citenamefont#1{#1}\fi
\expandafter\ifx\csname url\endcsname\relax
  \def\url#1{\texttt{#1}}\fi
\expandafter\ifx\csname urlprefix\endcsname\relax\def\urlprefix{URL }\fi
\providecommand{\bibinfo}[2]{#2}
\providecommand{\eprint}[2][]{\url{#2}}

\bibitem[{\citenamefont{Neto et~al.}(2009)\citenamefont{Neto, Guinea, Peres,
  Novoselov, and Geim}}]{neto09}
\bibinfo{author}{\bibfnamefont{A.~H.~C.} \bibnamefont{Neto}},
  \bibinfo{author}{\bibfnamefont{F.}~\bibnamefont{Guinea}},
  \bibinfo{author}{\bibfnamefont{N.~M.~R.} \bibnamefont{Peres}},
  \bibinfo{author}{\bibfnamefont{L.~S.} \bibnamefont{Novoselov}},
  \bibnamefont{and} \bibinfo{author}{\bibfnamefont{A.~K.} \bibnamefont{Geim}},
  \bibinfo{journal}{Rev. Mod. Phys.} \textbf{\bibinfo{volume}{81}},
  \bibinfo{pages}{109} (\bibinfo{year}{2009}).

\bibitem[{\citenamefont{Abergel et~al.}(2010)\citenamefont{Abergel, Alpakov,
  Berashevich, Ziegler, and Chakraborty}}]{abergel10}
\bibinfo{author}{\bibfnamefont{D.~S.~L.} \bibnamefont{Abergel}},
  \bibinfo{author}{\bibfnamefont{V.}~\bibnamefont{Alpakov}},
  \bibinfo{author}{\bibfnamefont{J.}~\bibnamefont{Berashevich}},
  \bibinfo{author}{\bibfnamefont{K.}~\bibnamefont{Ziegler}}, \bibnamefont{and}
  \bibinfo{author}{\bibfnamefont{T.}~\bibnamefont{Chakraborty}},
  \bibinfo{journal}{Adv. Phys.} \textbf{\bibinfo{volume}{59}},
  \bibinfo{pages}{261} (\bibinfo{year}{2010}).

\bibitem[{\citenamefont{Peres}(2010)}]{peres10}
\bibinfo{author}{\bibfnamefont{N.~M.~R.} \bibnamefont{Peres}},
  \bibinfo{journal}{Rev. Mod. Phys.} \textbf{\bibinfo{volume}{82}},
  \bibinfo{pages}{2673} (\bibinfo{year}{2010}).

\bibitem[{\citenamefont{Goerbig}(2011)}]{goerbig11}
\bibinfo{author}{\bibfnamefont{M.}~\bibnamefont{Goerbig}},
  \bibinfo{journal}{Rev. Mod. Phys.} \textbf{\bibinfo{volume}{83}},
  \bibinfo{pages}{1193} (\bibinfo{year}{2011}).

\bibitem[{\citenamefont{Kempa et~al.}(2006)\citenamefont{Kempa, Esquinazi, and
  Kopelevich}}]{kempa06}
\bibinfo{author}{\bibfnamefont{H.}~\bibnamefont{Kempa}},
  \bibinfo{author}{\bibfnamefont{P.}~\bibnamefont{Esquinazi}},
  \bibnamefont{and}
  \bibinfo{author}{\bibfnamefont{Y.}~\bibnamefont{Kopelevich}},
  \bibinfo{journal}{Solid State Commun.} \textbf{\bibinfo{volume}{138}},
  \bibinfo{pages}{118} (\bibinfo{year}{2006}).

\bibitem[{\citenamefont{Arovas and Guinea}(2008)}]{guinea08}
\bibinfo{author}{\bibfnamefont{D.~P.} \bibnamefont{Arovas}} \bibnamefont{and}
  \bibinfo{author}{\bibfnamefont{F.}~\bibnamefont{Guinea}},
  \bibinfo{journal}{Phys. Rev. B} \textbf{\bibinfo{volume}{78}},
  \bibinfo{pages}{245 416} (\bibinfo{year}{2008}).

\bibitem[{\citenamefont{Koshino and McCann}(2013)}]{koshino13}
\bibinfo{author}{\bibfnamefont{M.}~\bibnamefont{Koshino}} \bibnamefont{and}
  \bibinfo{author}{\bibfnamefont{E.}~\bibnamefont{McCann}},
  \bibinfo{journal}{Phys. Rev. B} \textbf{\bibinfo{volume}{87}},
  \bibinfo{pages}{45 420} (\bibinfo{year}{2013}).

\bibitem[{\citenamefont{Brandt et~al.}(1988)\citenamefont{Brandt, Chudinov, and
  Ponomarev}}]{brandt88}
\bibinfo{author}{\bibfnamefont{N.~B.} \bibnamefont{Brandt}},
  \bibinfo{author}{\bibfnamefont{S.~M.} \bibnamefont{Chudinov}},
  \bibnamefont{and} \bibinfo{author}{\bibfnamefont{Y.~G.}
  \bibnamefont{Ponomarev}}, \emph{\bibinfo{title}{Semimetals, 1. Graphite and
  its Compounds}} (\bibinfo{publisher}{North Holland}, \bibinfo{year}{1988}).

\bibitem[{\citenamefont{Grueneis et~al.}(2008)\citenamefont{Grueneis,
  Attaccalite, L.Wirtz, Shiozawa, Saito, Pichler, and Rubio}}]{grueneis08}
\bibinfo{author}{\bibfnamefont{A.}~\bibnamefont{Grueneis}},
  \bibinfo{author}{\bibfnamefont{C.}~\bibnamefont{Attaccalite}},
  \bibinfo{author}{\bibnamefont{L.Wirtz}},
  \bibinfo{author}{\bibfnamefont{H.}~\bibnamefont{Shiozawa}},
  \bibinfo{author}{\bibfnamefont{R.}~\bibnamefont{Saito}},
  \bibinfo{author}{\bibfnamefont{T.}~\bibnamefont{Pichler}}, \bibnamefont{and}
  \bibinfo{author}{\bibfnamefont{A.}~\bibnamefont{Rubio}},
  \bibinfo{journal}{Phys. Rev. B} \textbf{\bibinfo{volume}{78}},
  \bibinfo{pages}{205 425} (\bibinfo{year}{2008}).

\bibitem[{\citenamefont{Charlier et~al.}(1994)\citenamefont{Charlier, Gonze,
  and Michenaud}}]{gonze94}
\bibinfo{author}{\bibfnamefont{J.-C.} \bibnamefont{Charlier}},
  \bibinfo{author}{\bibfnamefont{X.}~\bibnamefont{Gonze}}, \bibnamefont{and}
  \bibinfo{author}{\bibfnamefont{J.-P.} \bibnamefont{Michenaud}},
  \bibinfo{journal}{Carbon} \textbf{\bibinfo{volume}{32}}, \bibinfo{pages}{289}
  (\bibinfo{year}{1994}).

\bibitem[{\citenamefont{Latil and Henrard}(2006)}]{latil06}
\bibinfo{author}{\bibfnamefont{S.}~\bibnamefont{Latil}} \bibnamefont{and}
  \bibinfo{author}{\bibfnamefont{L.}~\bibnamefont{Henrard}},
  \bibinfo{journal}{Phys. Rev. Lett.} \textbf{\bibinfo{volume}{97}},
  \bibinfo{pages}{036803} (\bibinfo{year}{2006}).

\bibitem[{\citenamefont{Aoki and Amawashi}(2007)}]{aoki07}
\bibinfo{author}{\bibfnamefont{M.}~\bibnamefont{Aoki}} \bibnamefont{and}
  \bibinfo{author}{\bibfnamefont{H.}~\bibnamefont{Amawashi}},
  \bibinfo{journal}{Solid State Commun.} \textbf{\bibinfo{volume}{142}},
  \bibinfo{pages}{123} (\bibinfo{year}{2007}).

\bibitem[{\citenamefont{Min et~al.}(2007)\citenamefont{Min, Sahu, Banerjee, and
  MacDonald}}]{min07}
\bibinfo{author}{\bibfnamefont{H.}~\bibnamefont{Min}},
  \bibinfo{author}{\bibfnamefont{B.}~\bibnamefont{Sahu}},
  \bibinfo{author}{\bibfnamefont{S.~K.} \bibnamefont{Banerjee}},
  \bibnamefont{and} \bibinfo{author}{\bibfnamefont{A.~H.}
  \bibnamefont{MacDonald}}, \bibinfo{journal}{Phys. Rev. B}
  \textbf{\bibinfo{volume}{75}}, \bibinfo{pages}{155115}
  (\bibinfo{year}{2007}).

\bibitem[{\citenamefont{Zhang et~al.}(2010)\citenamefont{Zhang, Sahu, Min, and
  MacDonald}}]{zhang10}
\bibinfo{author}{\bibfnamefont{F.}~\bibnamefont{Zhang}},
  \bibinfo{author}{\bibfnamefont{B.}~\bibnamefont{Sahu}},
  \bibinfo{author}{\bibfnamefont{H.}~\bibnamefont{Min}}, \bibnamefont{and}
  \bibinfo{author}{\bibfnamefont{A.~H.} \bibnamefont{MacDonald}},
  \bibinfo{journal}{Phys. Rev. B} \textbf{\bibinfo{volume}{82}},
  \bibinfo{pages}{35 409} (\bibinfo{year}{2010}).

\bibitem[{\citenamefont{Xiao et~al.}(2011)\citenamefont{Xiao, Tasnadi,
  Koepernik, Venderbos, Richter, and Taut}}]{xiao11}
\bibinfo{author}{\bibfnamefont{R.}~\bibnamefont{Xiao}},
  \bibinfo{author}{\bibfnamefont{F.}~\bibnamefont{Tasnadi}},
  \bibinfo{author}{\bibfnamefont{K.}~\bibnamefont{Koepernik}},
  \bibinfo{author}{\bibfnamefont{J.~W.~F.} \bibnamefont{Venderbos}},
  \bibinfo{author}{\bibfnamefont{M.}~\bibnamefont{Richter}}, \bibnamefont{and}
  \bibinfo{author}{\bibfnamefont{M.}~\bibnamefont{Taut}},
  \bibinfo{journal}{Phys. Rev. B} \textbf{\bibinfo{volume}{84}},
  \bibinfo{pages}{165 404} (\bibinfo{year}{2011}).

\bibitem[{fpl()}]{fplo}
\bibinfo{howpublished}{http://www.fplo.de/, version: 9.01-35-x86\_64}.

\bibitem[{\citenamefont{Koepernik and Eschrig}(1999)}]{koepernik99a}
\bibinfo{author}{\bibfnamefont{K.}~\bibnamefont{Koepernik}} \bibnamefont{and}
  \bibinfo{author}{\bibfnamefont{H.}~\bibnamefont{Eschrig}},
  \bibinfo{journal}{Phys. Rev. B} \textbf{\bibinfo{volume}{59}},
  \bibinfo{pages}{1743} (\bibinfo{year}{1999}).

\bibitem[{\citenamefont{Perdew and Wang}(1992)}]{PW92}
\bibinfo{author}{\bibfnamefont{J.}~\bibnamefont{Perdew}} \bibnamefont{and}
  \bibinfo{author}{\bibfnamefont{Y.}~\bibnamefont{Wang}},
  \bibinfo{journal}{Phys. Rev. B} \textbf{\bibinfo{volume}{45}},
  \bibinfo{pages}{13 244} (\bibinfo{year}{1992}).

\bibitem[{\citenamefont{Perdew et~al.}(1996)\citenamefont{Perdew, Burke, and
  Ernzerhof}}]{PBE96}
\bibinfo{author}{\bibfnamefont{J.}~\bibnamefont{Perdew}},
  \bibinfo{author}{\bibfnamefont{K.}~\bibnamefont{Burke}}, \bibnamefont{and}
  \bibinfo{author}{\bibfnamefont{M.}~\bibnamefont{Ernzerhof}},
  \bibinfo{journal}{Phys. Rev. Lett.} \textbf{\bibinfo{volume}{77}},
  \bibinfo{pages}{3 865} (\bibinfo{year}{1996}).

\bibitem[{\citenamefont{Ooi et~al.}(2006)\citenamefont{Ooi, Rairkar, and
  Adams}}]{ooi06}
\bibinfo{author}{\bibfnamefont{N.}~\bibnamefont{Ooi}},
  \bibinfo{author}{\bibfnamefont{A.}~\bibnamefont{Rairkar}}, \bibnamefont{and}
  \bibinfo{author}{\bibfnamefont{J.~B.} \bibnamefont{Adams}},
  \bibinfo{journal}{Carbon} \textbf{\bibinfo{volume}{44}}, \bibinfo{pages}{231}
  (\bibinfo{year}{2006}).

\bibitem[{\citenamefont{Lehmann and Taut}(1972)}]{tetrahedron1}
\bibinfo{author}{\bibfnamefont{G.}~\bibnamefont{Lehmann}} \bibnamefont{and}
  \bibinfo{author}{\bibfnamefont{M.}~\bibnamefont{Taut}},
  \bibinfo{journal}{phys. stat. sol. (b)} \textbf{\bibinfo{volume}{54}},
  \bibinfo{pages}{469} (\bibinfo{year}{1972}).

\bibitem[{\citenamefont{Lehmann and Taut}(1973)}]{tetrahedron2}
\bibinfo{author}{\bibfnamefont{G.}~\bibnamefont{Lehmann}} \bibnamefont{and}
  \bibinfo{author}{\bibfnamefont{M.}~\bibnamefont{Taut}},
  \bibinfo{journal}{phys. stat. sol. (b)} \textbf{\bibinfo{volume}{57}},
  \bibinfo{pages}{815} (\bibinfo{year}{1973}).

\bibitem[{\citenamefont{Kelly}(1981)}]{kelly81}
\bibinfo{author}{\bibfnamefont{B.~T.} \bibnamefont{Kelly}},
  \emph{\bibinfo{title}{Physics of Graphite}} (\bibinfo{publisher}{Applied
  Science Publishers}, \bibinfo{year}{1981}).

\bibitem[{\citenamefont{Charlier et~al.}(1991)\citenamefont{Charlier, Gonze,
  and Michenaud}}]{gonze91}
\bibinfo{author}{\bibfnamefont{J.-C.} \bibnamefont{Charlier}},
  \bibinfo{author}{\bibfnamefont{X.}~\bibnamefont{Gonze}}, \bibnamefont{and}
  \bibinfo{author}{\bibfnamefont{J.-P.} \bibnamefont{Michenaud}},
  \bibinfo{journal}{Phys. Rev. B} \textbf{\bibinfo{volume}{43}},
  \bibinfo{pages}{4579} (\bibinfo{year}{1991}).

\bibitem[{\citenamefont{Slonzewski and Weiss}(1958)}]{SWM1}
\bibinfo{author}{\bibfnamefont{J.~C.} \bibnamefont{Slonzewski}}
  \bibnamefont{and} \bibinfo{author}{\bibfnamefont{P.~R.} \bibnamefont{Weiss}},
  \bibinfo{journal}{Phys. Rev.} \textbf{\bibinfo{volume}{109}},
  \bibinfo{pages}{272} (\bibinfo{year}{1958}).

\bibitem[{\citenamefont{McClure}(1957)}]{SWM2}
\bibinfo{author}{\bibfnamefont{J.~W.} \bibnamefont{McClure}},
  \bibinfo{journal}{Phys. Rev.} \textbf{\bibinfo{volume}{108}},
  \bibinfo{pages}{612} (\bibinfo{year}{1957}).

\bibitem[{\citenamefont{Luk'yanchuk and Kopelevich}(2004)}]{kopelevich04}
\bibinfo{author}{\bibfnamefont{I.~A.} \bibnamefont{Luk'yanchuk}}
  \bibnamefont{and}
  \bibinfo{author}{\bibfnamefont{Y.}~\bibnamefont{Kopelevich}},
  \bibinfo{journal}{Phys. Rev. Lett.} \textbf{\bibinfo{volume}{93}},
  \bibinfo{pages}{166 402} (\bibinfo{year}{2004}).

\bibitem[{\citenamefont{Sugawara et~al.}(2006)\citenamefont{Sugawara, Sato,
  Souma, Takahashi, and Suematsu}}]{sugawa06}
\bibinfo{author}{\bibfnamefont{K.}~\bibnamefont{Sugawara}},
  \bibinfo{author}{\bibfnamefont{T.}~\bibnamefont{Sato}},
  \bibinfo{author}{\bibfnamefont{S.}~\bibnamefont{Souma}},
  \bibinfo{author}{\bibfnamefont{T.}~\bibnamefont{Takahashi}},
  \bibnamefont{and} \bibinfo{author}{\bibfnamefont{H.}~\bibnamefont{Suematsu}},
  \bibinfo{journal}{Phys. Rev. B} \textbf{\bibinfo{volume}{73}},
  \bibinfo{pages}{45 124} (\bibinfo{year}{2006}).

\bibitem[{\citenamefont{Heine}(1962)}]{heine62}
\bibinfo{author}{\bibfnamefont{V.}~\bibnamefont{Heine}},
  \bibinfo{journal}{Proc. Phys. Soc.} \textbf{\bibinfo{volume}{81}},
  \bibinfo{pages}{300} (\bibinfo{year}{1962}).

\bibitem[{not()}]{note1}
\bibinfo{howpublished}{Because the total energy differences between different
  stacking orders are exceptionally small, for the total energy calculations we
  used a refined real space integration mesh for the potential matrix elements.
  The default mesh produces noice in the 10$^{-5}$ eV energy range, which
  overshadows the physical energies in this case. We increased the number of
  radial mesh points to 200 and used the densest implemented angular mesh with
  602 points on the unit uphere for each radial shell.}

\bibitem[{\citenamefont{Grimme}(2011)}]{grimme11}
\bibinfo{author}{\bibfnamefont{S.}~\bibnamefont{Grimme}},
  \bibinfo{journal}{Comp. Mol. Sci.} \textbf{\bibinfo{volume}{1}},
  \bibinfo{pages}{211} (\bibinfo{year}{2011}).

\bibitem[{\citenamefont{Grimme}(2006)}]{grimme06}
\bibinfo{author}{\bibfnamefont{S.}~\bibnamefont{Grimme}}, \bibinfo{journal}{J.
  Comput. Chem.} \textbf{\bibinfo{volume}{27}}, \bibinfo{pages}{1 787}
  (\bibinfo{year}{2006}).

\bibitem[{\citenamefont{Tomanek et~al.}(1987)\citenamefont{Tomanek, Louie,
  Mamin, Abraham, Thomson, Ganz, and Clarke}}]{tomanek87}
\bibinfo{author}{\bibfnamefont{D.}~\bibnamefont{Tomanek}},
  \bibinfo{author}{\bibfnamefont{S.~G.} \bibnamefont{Louie}},
  \bibinfo{author}{\bibfnamefont{H.~J.} \bibnamefont{Mamin}},
  \bibinfo{author}{\bibfnamefont{D.~W.} \bibnamefont{Abraham}},
  \bibinfo{author}{\bibfnamefont{R.~E.} \bibnamefont{Thomson}},
  \bibinfo{author}{\bibfnamefont{E.}~\bibnamefont{Ganz}}, \bibnamefont{and}
  \bibinfo{author}{\bibfnamefont{J.}~\bibnamefont{Clarke}},
  \bibinfo{journal}{Phys. Rev. B} \textbf{\bibinfo{volume}{35}},
  \bibinfo{pages}{7790} (\bibinfo{year}{1987}).

\bibitem[{\citenamefont{Boettger}(1994)}]{boettger94}
\bibinfo{author}{\bibfnamefont{J.~C.} \bibnamefont{Boettger}},
  \bibinfo{journal}{Phys. Rev. B} \textbf{\bibinfo{volume}{49}},
  \bibinfo{pages}{16 798} (\bibinfo{year}{1994}).

\end{thebibliography}

\end{document}